\documentclass[review]{elsarticle}
\usepackage{appendix}
\usepackage{amsmath,amssymb}
\usepackage{gensymb}
\usepackage{blindtext}
\usepackage{booktabs}
\usepackage{multirow}
\begin{document}

\begin{frontmatter}

\title{Polarized Adding Method of Discrete Ordinate Approximation for Ultraviolet-Visible and Near-Infrared Radiative Transfer}

\author[a]{Kun Wu}
\author[b]{Feng Zhang\corref{mycorrespondingauthor}}
\cortext[mycorrespondingauthor]{Corresponding author}
\ead{fengzhang@fudan.edu.cn}
\author[b]{Wenwen Li}
\author[b]{Fengzi Bao}
\author[c]{Yi-ning Shi}

\address[a]{Key Laboratory of Meteorological Disaster, Ministry of Education (KLME)/ Joint International Research Laboratory of Climate and Environment Change (ILCEC)/ Collaborative Innovation Center on Forecast and Evaluation of Meteorological Disasters (CIC-FEMD) / Institute for Climate and Application Research (ICAR), Nanjing University of Information Science and Technology, Nanjing 210044, China}
\address[b]{Department of Atmospheric and Oceanic Sciences and Institutes of Atmospheric Sciences, Fudan University, Shanghai 200433, China}
\address[c]{CMA Earth System Modeling and Prediction Centre, China Meteorological Administration, Beijing 100081, China}

\begin{abstract}
The polarization characteristics of atmospheric scattering are important and should not be ignored in radiative transfer simulations. In this study, a new vector radiative transfer model called the polarized adding method of discrete ordinate approximation (POLDDA) is proposed for use in remote sensing applications for ultraviolet-visible and near-infrared spectra. The single-layer radiative transfer process and inhomogeneous multi-layer connection are solved using the discrete ordinate method (DOM) and adding methods, respectively. By combining the advantages of DOM and the adding method, the Stokes vector (including the $I$-, $Q$-, $U$-, and $V$-components) calculated using the new method conforms to the results of PolRadtran/RT3, whether in a Rayleigh scattering atmosphere or the water cloud case. Moreover, the relative root-mean-square error (RMSE) values of the Stokes vector for the test cases between MYSTIC and the new method or RT3 prove the accuracy of the proposed method. Meanwhile, the new method has a higher computational efficiency than RT3, particularly for an atmosphere with a large scattering optical depth. Unlike RT3, the computation time of the proposed method does not increase with the optical depth of each layer.
\end{abstract}


\end{frontmatter}

%\linenumbers

\section{Introduction}

In recent years, interest in the development of vector radiative transfer methods and instruments with polarization capabilities for aircraft, balloons, satellites, and ground-based platforms has increased rapidly. Radiation scattered by air molecules (Rayleigh scattering), aerosols, and cloud particles is polarized and shows different polarization characteristics based on the scattering event \citep{Duan2010, Yin2022}. The degree of linear and circular polarization produced by clouds and aerosol particles is more sensitive to the shape, size, and refractive index of the poly-dispersed small scattering particles. Polarization can provide additional information on atmospheric optical phenomena and their constituents \citep{Hansen1974}, including the properties of atmospheric components and vertical distributions. Numerous studies \citep{Mishchenko1994, Lacis1998, Oikarinen2001, Levy2004, Stam2005, Natraj2007} have shown that approximately 10\% of the errors introduced in atmospheric radiative transfer simulations and remote sensing are caused by ignoring the atmospheric polarization, particularly at short wavelengths \citep{Chandrasekhar1950, Lacis1998, Stam2005, Gao2010} owing to more scattering compared to long wavelengths \citep{Cai2023}. Therefore, a scalar radiative transfer model cannot sufficiently describe the nature of radiation processes \citep{Mishchenko2007}.

Various vector radiative transfer schemes have been developed, which commonly use the adding-doubling method \citep{Hansen1971, Plass1973, Haan1987, Evans1991, Bai2017, Bai2020}, discrete ordinates method \citep{Stamnes1984, Weng1992, Schulz1999, Siewert2000, Ota2010}, Monte-Carlo solutions method \cite{Kattawar1968, Collins1972, Wu1989, Roberti1999}, successive orders of scattering method \citep{Irvine1975, Min2004, Duan2010}, Invariant Imbedding Method \citep{Mishchenko1990}, and spherical harmonic method \citep{Evans1998}. Each method has its own advantages and disadvantages. For instance, the adding-doubling method deals with multiple scattering using the doubling method \citep{Evans1991}. Though the adding method can simplify the layer-to-layer connection, it is time-consuming to compute an atmospheric layer with a thick optical depth, such as a sky with thick stratus. This is caused by the increase in the number of thin sub-layers (a number 2$^{N}$ of identical thin layers), each of which is characterized by single scattering. The problem can be overcome using the discrete ordinate method (DOM). DOM solves the vector transfer equation using eigenvectors and eigenvalues \citep{Nakajima1986, Schulz1999}. It can calculate the interior variation of the reflection and transmission function within a layer; therefore, it is accurate and efficient for thick scattering media \citep{Weng1992}, such as aerosols and clouds.

In modeling scalar solar radiation, there is the method \citep{ZhangF2013} combines the advantages of the DOM and adding method, which via using DOM for a single-layer radiative transfer solution and the adding method employed for the inhomogeneous multi-layer connection in a plane-parallel scattering atmosphere. In this study, it will be expanded to vector radiative transfer to optimize the accuracy and computational efficiency. The remainder of this study is organized as follows. Section 2 describes the formulation and algorithm of the proposed method. In Section 3, the calculation accuracy and efficiency of the proposed method are evaluated for different scattering atmospheres. Finally, the conclusions and discussion are presented in Section 4.

\section{Solution by discrete ordinates method for a single-layer}
\subsection{Solar radiation}
The polarized radiative transfer equation with only a solar source in a plane-parallel medium is given by \citep{Chandrasekhar1950, Liu2002, Liou2005}
\begin{equation} \label{eq:1}
\begin{split}
\mu \frac{d\mathbf{L}(\tau,\mu,\varphi)}{d\tau} =& \mathbf{L}(\tau,\mu,\varphi) - \frac{\omega}{4\pi} \int_0 ^{2\pi} \int_{-1} ^1 \mathbf{Z}(\mu ,\varphi ,\mu ', \varphi ')\mathbf{L}(\tau,\mu',\varphi')d\mu' d\varphi' \\ & -\frac{\omega}{4\pi}\mathbf{Z}(\mu,\varphi,\mu_{0},\varphi_{0})\mathbf{F_{0}}(\mu_{0},\varphi_{0})e^{-\tau/\mu_{0}},
\end{split}
\end{equation}
where $\mathbf{L}=[I, Q, U , V]^\mathbb{T}$ represents the Stokes vector, and the superscript $\mathbb{T}$ indicates a matrix transpose. $\tau$ indicates the optical depth, $\omega$ signifies the single scattering albedo, $\varphi$ indicates the azimuthal angle, and $\mu$ denotes the zenith angle cosine.
$\mathbf{Z}$ is a 4$\times$4 scattering matrix, defined as
$\mathbf{Z}=\mathbf{C}(\pi-i_2)\mathbf{P}(\mu ,\varphi ,\mu ', \varphi ')\mathbf{C}(-i_1)$, where $\mathbf{C}$ represents the rotational matrices with $i_1$ and $i_2$ defining the rotation angles, and $\mathbf{P}$ represents the scattering phase matrix, and each element of $\mathbf{P}$ can be expanded using the Legendre series. $\mathbf{F_{0}}$ represents the incident solar Stokes vector at the top of the atmosphere (TOA).

The radiation transfer equation can be split into 2M equations by expanding the phase matrix and $\mathbf{L}$ into the Fourier cosine and sine series.
The discrete ordinate method is used to solve the equation and neglecting the superscript $m$, Eq.(1) can be written as in matrix form as
\begin{equation} \label{eq:2}
\frac{d}{d\tau}{\left[ {\begin{array}{*{20}{c}}
{{{\bf{L}}^ + }(\tau )}\\
{{{\bf{L}}^ - }(\tau )}
\end{array}} \right]} = {\left[
{\begin{array}{cc}
\bf{X}_{11} & \bf{X}_{12} \\
\bf{X}_{21} & \bf{X}_{22}
\end{array}} \right]}
{\left[ {\begin{array}{c}
{{{\bf{L}}^ + }}\\
{{{\bf{L}}^ - }}
\end{array}} \right]} +
{\left[ {\begin{array}{c}
\bf{E}^{+}\\
\bf{E}^{-}
\end{array}} \right]} e^{-\tau/\mu_{0}},
\end{equation}
where $\mathbf{L}^{\pm}(\tau) = [I(\tau,\pm \mu_{1}), \cdot \cdot \cdot,I(\tau,\pm \mu_{N}),Q(\tau,\pm \mu_{1}), \cdot  \cdot \cdot ,Q(\tau,\pm \mu_{N}), \\
U(\tau,\pm \mu_{1}), \cdot  \cdot  \cdot ,U(\tau,\pm \mu_{N}),V(\tau,\pm \mu_{1}), \cdot  \cdot  \cdot ,V(\tau,\pm \mu_{N})]^{\mathbb{T}}$.
The details are given in Appendix A.

This can be solved using the eigenvalue method \citep{Siewert2000} as follows:
\begin{equation} \label{eq:3}
{\left[ {\begin{array}{*{20}{c}}
{{{\bf{L}}^ + }(\tau )}\\
{{{\bf{L}}^ - }(\tau )}
\end{array}} \right]} = \bf{G}\circ\bf{K}{\left[
{\begin{array}{c}
\bf{C}_{1}  \\
\bf{C}_{2}
\end{array}} \right]} - \mu_{0}
{\left[ {\begin{array}{c}
{{{\bf{E}}^ + }}\\
{{{\bf{E}}^ - }}
\end{array}} \right]}e^{-\tau/\mu_{0}}
\end{equation}
where $\mathbf{G}$ is composed of the eigenvectors of $\mathbf{X}$, and $\mathbf{K}$ is composed of the eigenvalues of $\mathbf{X}$.
$[\bf{C}_{1},\bf{C}_{2}]^{\mathbb{T}}$ can be obtained from the boundary condition, that is, $\bf{L}^{-}(0)=0$ and $\bf{L}^{+}(\tau_{1})=0$. By solving Eq.(6), $\bf{L}^{+}(0)$ and $\bf{L}^{-}(\tau_{1})$ can be expressed as follows:
$\bf{L}^{+}(0)=\left[ {\begin{array}{c}
\mathbf{R}^{I}\\
\mathbf{R}^{Q}\\
\mathbf{R}^{U}\\
\mathbf{R}^{V}\\
\end{array}} \right]$,
$\bf{L}^{-}(\tau_{1})=\left[ {\begin{array}{*{20}{c}}
\mathbf{T}^{I}\\
\mathbf{T}^{Q}\\
\mathbf{T}^{U}\\
\mathbf{T}^{V}\\
\end{array}} \right]$.
$\mathbf{R}$ and $\mathbf{X}$ represent the direct reflection and transmission matrices, respectively, of a single layer with optical depth $\tau_{1}$, and can be expressed as
\begin{equation} \label{eq:4}
\begin{aligned}
\mathbf{R}^{I}=\left[
R^{I\leftarrow I}(0,\mu_{1},\mu_{0}),\cdots,R^{I\leftarrow I}(0,\mu_{N},\mu_{0})
\right]^{\mathbb{T}}, \\
\mathbf{R}^{Q}=\left[
R^{Q\leftarrow I}(0,\mu_{1},\mu_{0}),\cdots,R^{Q\leftarrow I}(0,\mu_{N},\mu_{0})
\right]^{\mathbb{T}}, \\
\mathbf{R}^{U}=\left[
R^{U\leftarrow I}(0,\mu_{1},\mu_{0}),\cdots,R^{U\leftarrow I}(0,\mu_{N},\mu_{0})
\right]^{\mathbb{T}}, \\
\mathbf{R}^{V}=\left[
R^{V\leftarrow I}(0,\mu_{1},\mu_{0}),\cdots,R^{V\leftarrow I}(0,\mu_{N},\mu_{0})
\right]^{\mathbb{T}}, \\
\mathbf{T}^{I}=\left[
T^{I\leftarrow I}(\tau_{1},\mu_{1},\mu_{0}),\cdots,T^{I\leftarrow I}(\tau_{1},\mu_{N},\mu_{0})
\right]^{\mathbb{T}}, \\
\mathbf{T}^{Q}=\left[
T^{Q\leftarrow I}(\tau_{1},\mu_{1},\mu_{0}),\cdots,T^{Q\leftarrow I}(\tau_{1},\mu_{N},\mu_{0})
\right]^{\mathbb{T}}, \\
\mathbf{T}^{U}=\left[
T^{U\leftarrow I}(\tau_{1},\mu_{1},\mu_{0}),\cdots,T^{U\leftarrow I}(\tau_{1},\mu_{N},\mu_{0})
\right]^{\mathbb{T}}, \\
\mathbf{T}^{V}=\left[
T^{V\leftarrow I}(\tau_{1},\mu_{1},\mu_{0}),\cdots,T^{V\leftarrow I}(\tau_{1},\mu_{N},\mu_{0})
\right]^{\mathbb{T}}.
\end{aligned}
\end{equation}
A connection exists between $I$-, $Q$-, $U$- and $V$-component transfer in a medium considering scattering. For the I-component, its change is not only from the I-component self, but also from the other three components. The superscript '$Q\leftarrow I$' denotes the reflection or transmission of $I$-component to $Q$-component dimension. The other parameters are defined in a similar manner.

\subsection{Diffuse radiation}
The diffuse equation of solar-polarized radiative transfer can be given as
\begin{equation} \label{eq:5}
\mu \frac{d\mathbf{L}(\tau,\mu,\varphi)}{d\tau} = \mathbf{L}(\tau,\mu,\varphi) - \frac{\omega}{4\pi} \int_0 ^{2\pi} \int_{-1} ^1 \mathbf{Z}(\mu ,\varphi ,\mu ', \varphi ')\mathbf{L}(\tau,\mu',\varphi')d\mu' d\varphi'.
\end{equation}
This solution is also required for the addition method. The details of the solutions of the diffuse radiation for the $I$ and $Q$ components are almost the same as those for the polarized thermal infrared radiative transfer described in \cite{Li2022}. The $U$ and $V$ components are obtained in a similar manner. Thus, the reflection and transmission matrices of diffuse radiation are given as
\begin{subequations} \label{eq:6}
\begin{align}
\overline{\mathbf{R}}&=\left[ {\begin{array}{cccc}
\overline {\cal R}^{I \leftarrow I}&{\overline {\cal R}}^{I \leftarrow Q}&{\overline {\cal R}}^{I \leftarrow U}&{\overline {\cal R}}^{I \leftarrow V}\\
{\overline {\cal R}}^{Q \leftarrow I}&{\overline {\cal R}}^{Q \leftarrow Q}&{\overline {\cal R}}^{Q \leftarrow U}&{\overline {\cal R}}^{Q \leftarrow V}\\
{\overline {\cal R}}^{U \leftarrow I}&{\overline {\cal R}}^{U \leftarrow Q}&{\overline {\cal R}}^{U \leftarrow U}&{\overline {\cal R}}^{U \leftarrow V}\\
{\overline {\cal R}}^{V \leftarrow I}&{\overline {\cal R}}^{V \leftarrow Q}&{\overline {\cal R}}^{V \leftarrow U}&{\overline {\cal R}}^{V \leftarrow V}\\
\end{array}} \right]_{4N\times4N} \\
\overline{\mathbf{T}}&=\left[ {\begin{array}{cccc}
{\overline {\cal T}}^{I \leftarrow I}\qquad{\overline {\cal T}}^{I \leftarrow Q}\qquad{\overline {\cal T}}^{I \leftarrow U}\qquad{\overline {\cal T}}^{I \leftarrow V}\\
{\overline {\cal T}}^{Q \leftarrow I}\qquad{\overline {\cal T}}^{Q \leftarrow Q}\qquad{\overline {\cal T}}^{Q \leftarrow U}\qquad{\overline {\cal T}}^{Q \leftarrow V}\\
{\overline {\cal T}}^{U \leftarrow I}\qquad{\overline {\cal T}}^{U \leftarrow Q}\qquad{\overline {\cal T}}^{U \leftarrow U}\qquad{\overline {\cal T}}^{U \leftarrow V}\\
{\overline {\cal T}}^{V \leftarrow I}\qquad{\overline {\cal T}}^{V \leftarrow Q}\qquad{\overline {\cal T}}^{V \leftarrow U}\qquad{\overline {\cal T}}^{V \leftarrow V}\\
\end{array}} \right]_{4N\times4N}.
\end{align}
\end{subequations}
where ${{\overline {\cal R}}^{I \leftarrow I}} = {\left[ {\begin{array}{*{20}{c}}
{(1+\delta_{0,m}){a_1}{R^{I \leftarrow I}}({\mu _1},{\mu _1}){\mu _1}}& \cdots &{(1+\delta_{0,m}){a_N}{R^{I \leftarrow I}}({\mu _1},{\mu _N}){\mu _N}}\\
 \cdots & \cdots & \cdots \\
{(1+\delta_{0,m}){a_1}{R^{I \leftarrow I}}({\mu _N},{\mu _1}){\mu _1}}& \cdots &{(1+\delta_{0,m}){a_N}{R^{I \leftarrow I}}({\mu _N},{\mu _N}){\mu _N}}
\end{array}} \right]_{N \times N}}$, \\
${{\overline {\cal T}}^{I \leftarrow I}} = {\left[ {\begin{array}{*{20}{c}}
{(1+\delta_{0,m}){a_1}{{\tilde T}^{I \leftarrow I}}({\mu _1},{\mu _1}){\mu _1}}& \cdots &{(1+\delta_{0,m}){a_N}{{\tilde T}^{I \leftarrow I}}({\mu _1},{\mu _N}){\mu _N}}\\
 \cdots & \cdots & \cdots \\
{(1+\delta_{0,m}){a_1}{{\tilde T}^{I \leftarrow I}}({\mu _N},{\mu _1}){\mu _1}}& \cdots &{(1+\delta_{0,m}){a_N}{{\tilde T}^{I \leftarrow I}}({\mu _N},{\mu _N}){\mu _N}}
\end{array}} \right]_{N \times N}}$, and other matrices on the right hand side of Eq.(6) are defined similarly.

\section{Four invariance principles and adding method for inhomogeneous multi-layer connection}
In this section, the four invariance principles \citep{Chandrasekhar1950} are extended to the polarized solar radiative transfer process and applied to the new model.
A schematic diagram is shown in Figure 1 (a)$\sim$(d): More specifically, we consider a combination of two inhomogeneous layers with $\tau_{1}$ (first layer) and $\tau_{2}$ (second layer) as the optical depths. $R_{i}^{a\leftarrow b}(\mu,\mu')$ and $T_{i}^{a\leftarrow b}(\mu,\mu')$ ($a=I,Q,U,V$ and $b=I,Q,U,V$) indicate the reflection and transmission functions of the $i$-th layer. The superscript $^*$ denotes that the radiation originates from below. From Figure 1, the four invariance principles can be expressed as follows:

(a) The reflected matrices $\mathbf{A}$ ($A^{I}(\mu)$ / $A^{Q}(\mu)$ / $A^{U}(\mu)$ / $A^{V}(\mu)$) of the Stokes vector at level 2 originate from two parts: the reflection of the direct solar beam by the second layer $\mathbf{R}_{2}(0,\mu)$ ($R^{I}_{2}(0,\mu)$ / $R^{Q}_{2}(0,\mu)$ / $R^{U}_{2}(0,\mu)$ / $R^{V}_{2}(0,\mu)$) and the coupled reflection of the downward components $\mathbf{D}$ ($D^{I}(\mu)$ / $D^{Q}(\mu)$ / $D^{U}(\mu)$  /  $D^{V}(\mu)$) through the second layer;

(b) The transmitted matrices $\mathbf{D}$ of the Stokes vector at level 2 originate from two parts: direct transmission by the first layer $\mathbf{T}_{1}(\tau_{1},-\mu)$ ($T^{I}_{1}(\tau_{1},-\mu)$ / $T^{Q}_{1}(\tau_{1},-\mu)$ / $T^{U}_{1}(\tau_{1},-\mu)$/ $T^{V}_{1}(\tau_{1},-\mu)$) and the coupled reflection of the upward components $\mathbf{A}$ through the first layer.

(c) The reflected matrices $\mathbf{R}_{1,2}(0,\mu)$ ($R^{I}_{1,2}(0,\mu)$ / $R^{Q}_{1,2}(0,\mu)$ / $R^{U}_{1,2}(0,\mu)$ / $R^{V}_{1,2}(0,\mu)$) of the Stokes vector at the top of the two-layer are composed of two parts: the reflection by the first layer $\mathbf{R}_{1}(0,\mu)$ ($R^{I}_{1}(0,\mu)$ / $R^{Q}_{1}(0,\mu)$ / $R^{U}_{1}(0,\mu)$ / $R^{V}_{1}(0,\mu)$), and the total transmission including the direct beam transmissions of $\exp(-\frac{\tau_{1}}{\mu})$ and $\mathbf{T}_{1}^{*}(\mu,\mu')$ of the upward components $\mathbf{A}$ through the first layer.

(d) The transmitted matrices $\mathbf{T}_{1,2}(\tau_{1}+\tau_{2},-\mu)$ ($T^{I}_{1,2}(\tau_{1}+\tau_{2},-\mu)$ / $T^{Q}_{1,2}(\tau_{1}+\tau_{2},-\mu)$ / $T^{U}_{1,2}(\tau_{1}+\tau_{2},-\mu)$ / $T^{V}_{1,2}(\tau_{1}+\tau_{2},-\mu)$) of the Stokes vector at the bottom of the two-layer are composed of two parts: transmission of the direct solar beam $\exp(-\frac{\tau_{1}}{\mu_{0}})$  through the second layer and the total coupled transmission, including $\exp(-\frac{\tau_{2}}{\mu})$ and $\mathbf{T}_{2}(\mu,\mu')$ of the downward components $\mathbf{D}$ through the second layer.

Based on the above (a$\sim$d) statements, we can write the four invariance principles in vector solar radiative transfer as
\begin{subequations}\label{eq:7}
\begin{align}
&\mathbf{A}=\mathbf{R}_{2}e^{-\frac{\tau_{1}}{\mu_{0}}} +
2\int_{0}^{1}\overline{\mathbf{R}}_{2} \mathbf{D}\mu'd\mu',\\
&\mathbf{D}=\mathbf{T}_{1}+2\int_{0}^{1}\overline{\mathbf{R}}^{*}_{1}\mathbf{A}\mu'd\mu',\\
&\mathbf{R}_{1,2}=\mathbf{R}_{1}+ \mathbf{A}e^{-\frac{\tau_{1}}{\mu_{0}}} + 2\int_{0}^{1}\overline{\mathbf{T}}^{*}_{1}\mathbf{A}\mu'd\mu',\\
&\mathbf{T}_{1,2}=\mathbf{T}_{2}e^{-\frac{\tau_{1}}{\mu_{0}}}+ \mathbf{D}e^{-\frac{\tau_{2}}{\mu_{0}}} + 2\int_{0}^{1}\overline{\mathbf{T}}_{2} \mathbf{D}\mu'd\mu'.
\end{align}
\end{subequations}
In Eq.(7),
$\mathbf{A}=[\mathbf{A}^{I},\mathbf{A}^{Q},\mathbf{A}^{U},\mathbf{A}^{V}]^\mathbb{T} $ ,
$\mathbf{D}=[\mathbf{D}^{I},\mathbf{D}^{Q},\mathbf{D}^{U},\mathbf{D}^{V}]^\mathbb{T} $ , \\
$\mathbf{R}_{1,2}=[\mathbf{R}_{1,2}^{I},\mathbf{R}_{1,2}^{Q},\mathbf{R}_{1,2}^{U},\mathbf{R}_{1,2}^{V}]^\mathbb{T}$,
$\mathbf{T}_{1,2}=[\mathbf{T}_{1,2}^{I},\mathbf{T}_{1,2}^{Q},\mathbf{T}_{1,2}^{U},\mathbf{T}_{1,2}^{V}]^\mathbb{T}$.\\
Others matrix in Eq.(7) are detailed in the Appendix B.

$N$-node Gaussian integration was used to handle the integration; thus, Eq.(7) can be written as follows:
\begin{subequations}\label{eq:8}
\begin{align}
\left[
\begin{array} {cccc}
\mathbf{A}^{I}  \\
\mathbf{A}^{Q}  \\
\mathbf{A}^{U}  \\
\mathbf{A}^{V} \\
\end{array}
\right]=\left[
\begin{array} {cccc}
\mathbf{R}_{2}^{I}  \\
\mathbf{R}_{2}^{Q}  \\
\mathbf{R}_{2}^{U}  \\
\mathbf{R}_{2}^{V}  \\
\end{array}
\right]e^{-\frac{\tau_{1}}{\mu_{0}}} + \left[
\begin{array} {cccc}
\overline{\cal R}_{2}^{I\leftarrow I} & \overline{\cal R}_{2}^{I\leftarrow Q} & \overline{\cal R}_{2}^{I\leftarrow U} & \overline{\cal R}_{2}^{I\leftarrow V} \\
\overline{\cal R}_{2}^{Q\leftarrow I} & \overline{\cal R}_{2}^{Q\leftarrow Q} & \overline{\cal R}_{2}^{Q\leftarrow U} & \overline{\cal R}_{2}^{Q\leftarrow V} \\
\overline{\cal R}_{2}^{U\leftarrow I} & \overline{\cal R}_{2}^{U\leftarrow Q} & \overline{\cal R}_{2}^{U\leftarrow U} & \overline{\cal R}_{2}^{U\leftarrow V} \\
\overline{\cal R}_{2}^{V\leftarrow I} & \overline{\cal R}_{2}^{V\leftarrow Q} & \overline{\cal R}_{2}^{V\leftarrow U} & \overline{\cal R}_{2}^{V\leftarrow V} \\
\end{array}
\right]
\left[
\begin{array} {cccc}
\mathbf{D}^{I}  \\
\mathbf{D}^{Q}  \\
\mathbf{D}^{U}  \\
\mathbf{D}^{V} \\
\end{array}
\right] \\
\left[
\begin{array} {cccc}
\mathbf{D}^{I}  \\
\mathbf{D}^{Q}  \\
\mathbf{D}^{U}  \\
\mathbf{D}^{V} \\
\end{array}
\right] = \left[
\begin{array} {cccc}
\mathbf{T}_{1}^{I}  \\
\mathbf{T}_{1}^{Q}  \\
\mathbf{T}_{1}^{U}  \\
\mathbf{T}_{1}^{V}  \\
\end{array}
\right] + \left[
\begin{array} {cccc}
\overline{\cal R}_{1}^{*,I\leftarrow I} & \overline{\cal R}_{1}^{*,I\leftarrow Q} & \overline{\cal R}_{1}^{*,I\leftarrow U} & \overline{\cal R}_{1}^{*,I\leftarrow V} \\
\overline{\cal R}_{1}^{*,Q\leftarrow I} & \overline{\cal R}_{1}^{*,Q\leftarrow Q} & \overline{\cal R}_{1}^{*,Q\leftarrow U} & \overline{\cal R}_{1}^{*,Q\leftarrow V} \\
\overline{\cal R}_{1}^{*,U\leftarrow I} & \overline{\cal R}_{1}^{*,U\leftarrow Q} & \overline{\cal R}_{1}^{*,U\leftarrow U} & \overline{\cal R}_{1}^{*,U\leftarrow V} \\
\overline{\cal R}_{1}^{*,V\leftarrow I} & \overline{\cal R}_{1}^{*,V\leftarrow Q} & \overline{\cal R}_{1}^{*,V\leftarrow U} & \overline{\cal R}_{1}^{*,V\leftarrow V} \\
\end{array}
\right]
\left[
\begin{array} {cccc}
\mathbf{A}^{I}  \\
\mathbf{A}^{Q}  \\
\mathbf{A}^{U}  \\
\mathbf{A}^{V} \\
\end{array}
\right] \\
\left[
\begin{array} {cccc}
\mathbf{R}^{I}_{1,2}  \\
\mathbf{R}^{Q}_{1,2}  \\
\mathbf{R}^{U}_{1,2}  \\
\mathbf{R}^{V}_{1,2} \\
\end{array}
\right]=\left[
\begin{array} {cccc}
\mathbf{R}_{1}^{I}  \\
\mathbf{R}_{1}^{Q}  \\
\mathbf{R}_{1}^{U}  \\
\mathbf{R}_{1}^{V}  \\
\end{array}
\right] + \left[
\begin{array} {cccc}
\overline {\cal T}_{1}^{*,I\leftarrow I} & \overline {\cal T}_{1}^{*,I\leftarrow Q} & \overline {\cal T}_{1}^{*,I\leftarrow U} & \overline {\cal T}_{1}^{*,I\leftarrow V} \\
\overline {\cal T}_{1}^{*,Q\leftarrow I} & \overline {\cal T}_{1}^{*,Q\leftarrow Q} & \overline {\cal T}_{1}^{*,Q\leftarrow U} & \overline {\cal T}_{1}^{*,Q\leftarrow V} \\
\overline {\cal T}_{1}^{*,U\leftarrow I} & \overline {\cal T}_{1}^{*,U\leftarrow Q} & \overline {\cal T}_{1}^{*,U\leftarrow U} & \overline {\cal T}_{1}^{*,U\leftarrow V} \\
\overline {\cal T}_{1}^{*,V\leftarrow I} & \overline {\cal T}_{1}^{*,V\leftarrow Q} & \overline {\cal T}_{1}^{*,V\leftarrow U} & \overline {\cal T}_{1}^{*,V\leftarrow V} \\
\end{array}
\right]
\left[
\begin{array} {cccc}
\mathbf{A}^{I}  \\
\mathbf{A}^{Q}  \\
\mathbf{A}^{U}  \\
\mathbf{A}^{V} \\
\end{array}
\right] \\
\left[
\begin{array} {cccc}
\mathbf{T}^{I}_{1,2}  \\
\mathbf{T}^{Q}_{1,2}  \\
\mathbf{T}^{U}_{1,2}  \\
\mathbf{T}^{V}_{1,2} \\
\end{array}
\right]=\left[
\begin{array} {cccc}
\mathbf{T}_{2}^{I}  \\
\mathbf{T}_{2}^{Q}  \\
\mathbf{T}_{2}^{U}  \\
\mathbf{T}_{2}^{V}  \\
\end{array}
\right]e^{-\frac{\tau_{1}}{\mu_{0}}} + \left[
\begin{array} {cccc}
\overline {\cal T}_{2}^{I\leftarrow I} & \overline {\cal T}_{2}^{I\leftarrow Q} & \overline {\cal T}_{2}^{I\leftarrow U} & \overline {\cal T}_{2}^{I\leftarrow V} \\
\overline {\cal T}_{2}^{Q\leftarrow I} & \overline {\cal T}_{2}^{Q\leftarrow Q} & \overline {\cal T}_{2}^{Q\leftarrow U} & \overline {\cal T}_{2}^{Q\leftarrow V} \\
\overline {\cal T}_{2}^{U\leftarrow I} & \overline {\cal T}_{2}^{U\leftarrow Q} & \overline {\cal T}_{2}^{U\leftarrow U} & \overline {\cal T}_{2}^{U\leftarrow V} \\
\overline {\cal T}_{2}^{V\leftarrow I} & \overline {\cal T}_{2}^{V\leftarrow Q} & \overline {\cal T}_{2}^{V\leftarrow U} & \overline {\cal T}_{2}^{V\leftarrow V} \\
\end{array}
\right]
\left[
\begin{array} {cccc}
\mathbf{D}^{I}  \\
\mathbf{D}^{Q}  \\
\mathbf{D}^{U}  \\
\mathbf{D}^{V} \\
\end{array}
\right]
\end{align}
\end{subequations}

From (8a-b), we obtain
\begin{subequations}\label{eq:9}
\begin{align}
&\left[
\begin{array} {cccc}
\mathbf{A}^{I}  \\
\mathbf{A}^{Q}  \\
\mathbf{A}^{U}  \\
\mathbf{A}^{V} \\
\end{array}
\right]=\overline{\mathbf{X}}_{2}^{1}\mathbf{Y}_{2}^{1}, \\
&\left[
\begin{array} {cccc}
\mathbf{D}^{I}  \\
\mathbf{D}^{Q}  \\
\mathbf{D}^{U}  \\
\mathbf{D}^{V} \\
\end{array}
\right]=\left[
\begin{array} {cccc}
\mathbf{T}_{1}^{I}  \\
\mathbf{T}_{1}^{Q}  \\
\mathbf{T}_{1}^{U}  \\
\mathbf{T}_{1}^{V}  \\
\end{array}
\right]+\left[
\begin{array} {cccc}
\overline{\cal R}_{1}^{*,I\leftarrow I} & \overline{\cal R}_{1}^{*,I\leftarrow Q} & \overline{\cal R}_{1}^{*,I\leftarrow U} & \overline{\cal R}_{1}^{*,I\leftarrow V} \\
\overline{\cal R}_{1}^{*,Q\leftarrow I} & \overline{\cal R}_{1}^{*,Q\leftarrow Q} & \overline{\cal R}_{1}^{*,Q\leftarrow U} & \overline{\cal R}_{1}^{*,Q\leftarrow V} \\
\overline{\cal R}_{1}^{*,U\leftarrow I} & \overline{\cal R}_{1}^{*,U\leftarrow Q} & \overline{\cal R}_{1}^{*,U\leftarrow U} & \overline{\cal R}_{1}^{*,U\leftarrow V} \\
\overline{\cal R}_{1}^{*,V\leftarrow I} & \overline{\cal R}_{1}^{*,V\leftarrow Q} & \overline{\cal R}_{1}^{*,V\leftarrow U} & \overline{\cal R}_{1}^{*,V\leftarrow V} \\
\end{array}
\right]
\overline{\mathbf{X}}_{2}^{1}\mathbf{Y}_{2}^{1}
\end{align}
\end{subequations}
where $\overline{\mathbf{X}}_{i}^{j}$ and $\mathbf{Y}_{i}^{j}$ are given in Appendix B.

Substituting (9) into (8c-d), the direct reflection and transmission can be expressed as follows:
\begin{subequations}\label{eq:10}
\begin{align}
&\left[
\begin{array} {cccc}
\mathbf{R}^{I}_{1,2}  \\
\mathbf{R}^{Q}_{1,2}  \\
\mathbf{R}^{U}_{1,2}  \\
\mathbf{R}^{V}_{1,2} \\
\end{array}
\right]=\left[
\begin{array} {cccc}
\mathbf{R}_{1}^{I}  \\
\mathbf{R}_{1}^{Q}  \\
\mathbf{R}_{1}^{U}  \\
\mathbf{R}_{1}^{V}  \\
\end{array}
\right] + \left[
\begin{array} {cccc}
\overline {\cal T}_{1}^{*,I\leftarrow I} & \overline {\cal T}_{1}^{*,I\leftarrow Q} & \overline {\cal T}_{1}^{*,I\leftarrow U} & \overline {\cal T}_{1}^{*,I\leftarrow V} \\
\overline {\cal T}_{1}^{*,Q\leftarrow I} & \overline {\cal T}_{1}^{*,Q\leftarrow Q} & \overline {\cal T}_{1}^{*,Q\leftarrow U} & \overline {\cal T}_{1}^{*,Q\leftarrow V} \\
\overline {\cal T}_{1}^{*,U\leftarrow I} & \overline {\cal T}_{1}^{*,U\leftarrow Q} & \overline {\cal T}_{1}^{*,U\leftarrow U} & \overline {\cal T}_{1}^{*,U\leftarrow V} \\
\overline {\cal T}_{1}^{*,V\leftarrow I} & \overline {\cal T}_{1}^{*,V\leftarrow Q} & \overline {\cal T}_{1}^{*,V\leftarrow U} & \overline {\cal T}_{1}^{*,V\leftarrow V} \\
\end{array}
\right]\overline{\mathbf{X}}_{2}^{1}\mathbf{Y}_{2}^{1},\\ \nonumber
& \left[
\begin{array} {cccc}
\mathbf{T}^{I}_{1,2}  \\
\mathbf{T}^{Q}_{1,2}  \\
\mathbf{T}^{U}_{1,2}  \\
\mathbf{T}^{V}_{1,2} \\
\end{array}
\right]=\left[
\begin{array} {cccc}
\mathbf{T}_{2}^{I}  \\
\mathbf{T}_{2}^{Q}  \\
\mathbf{T}_{2}^{U}  \\
\mathbf{T}_{2}^{V}  \\
\end{array}
\right]e^{-\frac{\tau_{1}}{\mu_{0}}} + \left[
\begin{array} {cccc}
\overline {\cal T}_{2}^{I\leftarrow I} & \overline {\cal T}_{2}^{I\leftarrow Q} & \overline {\cal T}_{2}^{I\leftarrow U} & \overline {\cal T}_{2}^{I\leftarrow V} \\
\overline {\cal T}_{2}^{Q\leftarrow I} & \overline {\cal T}_{2}^{Q\leftarrow Q} & \overline {\cal T}_{2}^{Q\leftarrow U} & \overline {\cal T}_{2}^{Q\leftarrow V} \\
\overline {\cal T}_{2}^{U\leftarrow I} & \overline {\cal T}_{2}^{U\leftarrow Q} & \overline {\cal T}_{2}^{U\leftarrow U} & \overline {\cal T}_{2}^{U\leftarrow V} \\
\overline {\cal T}_{2}^{V\leftarrow I} & \overline {\cal T}_{2}^{V\leftarrow Q} & \overline {\cal T}_{2}^{V\leftarrow U} & \overline {\cal T}_{2}^{V\leftarrow V} \\
\end{array}
\right] \left[
\begin{array} {cccc}
\mathbf{T}_{1}^{I}  \\
\mathbf{T}_{1}^{Q}  \\
\mathbf{T}_{1}^{U}  \\
\mathbf{T}_{1}^{V}  \\
\end{array}
\right] + \\
&\quad \left[
\begin{array} {cccc}
\overline {\cal T}_{2}^{I\leftarrow I} & \overline {\cal T}_{2}^{I\leftarrow Q} & \overline {\cal T}_{2}^{I\leftarrow U} & \overline {\cal T}_{2}^{I\leftarrow V} \\
\overline {\cal T}_{2}^{Q\leftarrow I} & \overline {\cal T}_{2}^{Q\leftarrow Q} & \overline {\cal T}_{2}^{Q\leftarrow U} & \overline {\cal T}_{2}^{Q\leftarrow V} \\
\overline {\cal T}_{2}^{U\leftarrow I} & \overline {\cal T}_{2}^{U\leftarrow Q} & \overline {\cal T}_{2}^{U\leftarrow U} & \overline {\cal T}_{2}^{U\leftarrow V} \\
\overline {\cal T}_{2}^{V\leftarrow I} & \overline {\cal T}_{2}^{V\leftarrow Q} & \overline {\cal T}_{2}^{V\leftarrow U} & \overline {\cal T}_{2}^{V\leftarrow V} \\
\end{array}
\right]
\left[
\begin{array} {cccc}
\overline{\cal R}_{1}^{*,I\leftarrow I} & \overline{\cal R}_{1}^{*,I\leftarrow Q} & \overline{\cal R}_{1}^{*,I\leftarrow U} & \overline{\cal R}_{1}^{*,I\leftarrow V} \\
\overline{\cal R}_{1}^{*,Q\leftarrow I} & \overline{\cal R}_{1}^{*,Q\leftarrow Q} & \overline{\cal R}_{1}^{*,Q\leftarrow U} & \overline{\cal R}_{1}^{*,Q\leftarrow V} \\
\overline{\cal R}_{1}^{*,U\leftarrow I} & \overline{\cal R}_{1}^{*,U\leftarrow Q} & \overline{\cal R}_{1}^{*,U\leftarrow U} & \overline{\cal R}_{1}^{*,U\leftarrow V} \\
\overline{\cal R}_{1}^{*,V\leftarrow I} & \overline{\cal R}_{1}^{*,V\leftarrow Q} & \overline{\cal R}_{1}^{*,V\leftarrow U} & \overline{\cal R}_{1}^{*,V\leftarrow V} \\
\end{array}
\right] \overline{\mathbf{X}}_{2}^{1}\mathbf{Y}_{2}^{1}
\end{align}
\end{subequations}

By applying the invariance principles to diffuse radiation, we obtain
\begin{subequations}\label{eq:11}
\begin{align}
&\left[
\begin{array} {cccc}
\overline{\cal R}_{1,2}^{I\leftarrow I} & \overline{\cal R}_{1,2}^{I\leftarrow Q} & \overline{\cal R}_{1,2}^{I\leftarrow U} & \overline{\cal R}_{1,2}^{I\leftarrow V} \\
\overline{\cal R}_{1,2}^{Q\leftarrow I} & \overline{\cal R}_{1,2}^{Q\leftarrow Q} & \overline{\cal R}_{1,2}^{Q\leftarrow U} & \overline{\cal R}_{1,2}^{Q\leftarrow V} \\
\overline{\cal R}_{1,2}^{U\leftarrow I} & \overline{\cal R}_{1,2}^{U\leftarrow Q} & \overline{\cal R}_{1,2}^{U\leftarrow U} & \overline{\cal R}_{1,2}^{U\leftarrow V} \\
\overline{\cal R}_{1,2}^{V\leftarrow I} & \overline{\cal R}_{1,2}^{V\leftarrow Q} & \overline{\cal R}_{1,2}^{V\leftarrow U} & \overline{\cal R}_{1,2}^{V\leftarrow V} \\
\end{array}
\right] =\left[
\begin{array} {cccc}
\overline{\cal R}_{1}^{I\leftarrow I} & \overline{\cal R}_{1}^{I\leftarrow Q} & \overline{\cal R}_{1}^{I\leftarrow U} & \overline{\cal R}_{1}^{I\leftarrow V} \\
\overline{\cal R}_{1}^{Q\leftarrow I} & \overline{\cal R}_{1}^{Q\leftarrow Q} & \overline{\cal R}_{1}^{Q\leftarrow U} & \overline{\cal R}_{1}^{Q\leftarrow V} \\
\overline{\cal R}_{1}^{U\leftarrow I} & \overline{\cal R}_{1}^{U\leftarrow Q} & \overline{\cal R}_{1}^{U\leftarrow U} & \overline{\cal R}_{1}^{U\leftarrow V} \\
\overline{\cal R}_{1}^{V\leftarrow I} & \overline{\cal R}_{1}^{V\leftarrow Q} & \overline{\cal R}_{1}^{V\leftarrow U} & \overline{\cal R}_{1}^{V\leftarrow V} \\
\end{array}
\right] \nonumber \\
& \quad + \left[
\begin{array} {cccc}
\overline {\cal T}_{1}^{*,I\leftarrow I} & \overline {\cal T}_{1}^{*,I\leftarrow Q} & \overline {\cal T}_{1}^{*,I\leftarrow U} & \overline {\cal T}_{1}^{*,I\leftarrow V} \\
\overline {\cal T}_{1}^{*,Q\leftarrow I} & \overline {\cal T}_{1}^{*,Q\leftarrow Q} & \overline {\cal T}_{1}^{*,Q\leftarrow U} & \overline {\cal T}_{1}^{*,Q\leftarrow V} \\
\overline {\cal T}_{1}^{*,U\leftarrow I} & \overline {\cal T}_{1}^{*,U\leftarrow Q} & \overline {\cal T}_{1}^{*,U\leftarrow U} & \overline {\cal T}_{1}^{*,U\leftarrow V} \\
\overline {\cal T}_{1}^{*,V\leftarrow I} & \overline {\cal T}_{1}^{*,V\leftarrow Q} & \overline {\cal T}_{1}^{*,V\leftarrow U} & \overline {\cal T}_{1}^{*,V\leftarrow V} \\
\end{array}
\right]\overline{\mathbf{X}}_{2}^{1} \nonumber \\
&\quad \times\left[
\begin{array} {cccc}
\overline{\cal R}_{2}^{I\leftarrow I} & \overline{\cal R}_{2}^{I\leftarrow Q} & \overline{\cal R}_{2}^{I\leftarrow U} & \overline{\cal R}_{2}^{I\leftarrow V} \\
\overline{\cal R}_{2}^{Q\leftarrow I} & \overline{\cal R}_{2}^{Q\leftarrow Q} & \overline{\cal R}_{2}^{Q\leftarrow U} & \overline{\cal R}_{2}^{Q\leftarrow V} \\
\overline{\cal R}_{2}^{U\leftarrow I} & \overline{\cal R}_{2}^{U\leftarrow Q} & \overline{\cal R}_{2}^{U\leftarrow U} & \overline{\cal R}_{2}^{U\leftarrow V} \\
\overline{\cal R}_{2}^{V\leftarrow I} & \overline{\cal R}_{2}^{V\leftarrow Q} & \overline{\cal R}_{2}^{V\leftarrow U} & \overline{\cal R}_{2}^{V\leftarrow V} \\
\end{array}
\right]
\left[
\begin{array} {cccc}
\overline {\cal T}_{1}^{I\leftarrow I} & \overline {\cal T}_{1}^{I\leftarrow Q} & \overline {\cal T}_{1}^{I\leftarrow U} & \overline {\cal T}_{1}^{I\leftarrow V} \\
\overline {\cal T}_{1}^{Q\leftarrow I} & \overline {\cal T}_{1}^{Q\leftarrow Q} & \overline {\cal T}_{1}^{Q\leftarrow U} & \overline {\cal T}_{1}^{Q\leftarrow V} \\
\overline {\cal T}_{1}^{U\leftarrow I} & \overline {\cal T}_{1}^{U\leftarrow Q} & \overline {\cal T}_{1}^{U\leftarrow U} & \overline {\cal T}_{1}^{U\leftarrow V} \\
\overline {\cal T}_{1}^{V\leftarrow I} & \overline {\cal T}_{1}^{V\leftarrow Q} & \overline {\cal T}_{1}^{V\leftarrow U} & \overline {\cal T}_{1}^{V\leftarrow V} \\
\end{array}
\right],  \\
&\left[
\begin{array} {cccc}
\overline {\cal T}_{1,2}^{I\leftarrow I} & \overline {\cal T}_{1,2}^{I\leftarrow Q} & \overline {\cal T}_{1,2}^{I\leftarrow U} & \overline {\cal T}_{1,2}^{I\leftarrow V} \\
\overline {\cal T}_{1,2}^{Q\leftarrow I} & \overline {\cal T}_{1,2}^{Q\leftarrow Q} & \overline {\cal T}_{1,2}^{Q\leftarrow U} & \overline {\cal T}_{1,2}^{Q\leftarrow V} \\
\overline {\cal T}_{1,2}^{U\leftarrow I} & \overline {\cal T}_{1,2}^{U\leftarrow Q} & \overline {\cal T}_{1,2}^{U\leftarrow U} & \overline {\cal T}_{1,2}^{U\leftarrow V} \\
\overline {\cal T}_{1,2}^{V\leftarrow I} & \overline {\cal T}_{1,2}^{V\leftarrow Q} & \overline {\cal T}_{1,2}^{V\leftarrow U} & \overline {\cal T}_{1,2}^{V\leftarrow V} \\
\end{array}
\right]=\left[
\begin{array} {cccc}
\overline {\cal T}_{2}^{I\leftarrow I} & \overline {\cal T}_{2}^{I\leftarrow Q} & \overline {\cal T}_{2}^{I\leftarrow U} & \overline {\cal T}_{2}^{I\leftarrow V} \\
\overline {\cal T}_{2}^{Q\leftarrow I} & \overline {\cal T}_{2}^{Q\leftarrow Q} & \overline {\cal T}_{2}^{Q\leftarrow U} & \overline {\cal T}_{2}^{Q\leftarrow V} \\
\overline {\cal T}_{2}^{U\leftarrow I} & \overline {\cal T}_{2}^{U\leftarrow Q} & \overline {\cal T}_{2}^{U\leftarrow U} & \overline {\cal T}_{2}^{U\leftarrow V} \\
\overline {\cal T}_{2}^{V\leftarrow I} & \overline {\cal T}_{2}^{V\leftarrow Q} & \overline {\cal T}_{2}^{V\leftarrow U} & \overline {\cal T}_{2}^{V\leftarrow V} \\
\end{array}
\right] \nonumber \\
&\quad \times \left[
\begin{array} {cccc}
\overline {\cal T}_{1}^{I\leftarrow I} & \overline {\cal T}_{1}^{I\leftarrow Q} & \overline {\cal T}_{1}^{I\leftarrow U} & \overline {\cal T}_{1}^{I\leftarrow V} \\
\overline {\cal T}_{1}^{Q\leftarrow I} & \overline {\cal T}_{1}^{Q\leftarrow Q} & \overline {\cal T}_{1}^{Q\leftarrow U} & \overline {\cal T}_{1}^{Q\leftarrow V} \\
\overline {\cal T}_{1}^{U\leftarrow I} & \overline {\cal T}_{1}^{U\leftarrow Q} & \overline {\cal T}_{1}^{U\leftarrow U} & \overline {\cal T}_{1}^{U\leftarrow V} \\
\overline {\cal T}_{1}^{V\leftarrow I} & \overline {\cal T}_{1}^{V\leftarrow Q} & \overline {\cal T}_{1}^{V\leftarrow U} & \overline {\cal T}_{1}^{V\leftarrow V} \\
\end{array}
\right]  + \left[
\begin{array} {cccc}
\overline {\cal T}_{2}^{I\leftarrow I} & \overline {\cal T}_{2}^{I\leftarrow Q} & \overline {\cal T}_{2}^{I\leftarrow U} & \overline {\cal T}_{2}^{I\leftarrow V} \\
\overline {\cal T}_{2}^{Q\leftarrow I} & \overline {\cal T}_{2}^{Q\leftarrow Q} & \overline {\cal T}_{2}^{Q\leftarrow U} & \overline {\cal T}_{2}^{Q\leftarrow V} \\
\overline {\cal T}_{2}^{U\leftarrow I} & \overline {\cal T}_{2}^{U\leftarrow Q} & \overline {\cal T}_{2}^{U\leftarrow U} & \overline {\cal T}_{2}^{U\leftarrow V} \\
\overline {\cal T}_{2}^{V\leftarrow I} & \overline {\cal T}_{2}^{V\leftarrow Q} & \overline {\cal T}_{2}^{V\leftarrow U} & \overline {\cal T}_{2}^{V\leftarrow V} \\
\end{array}
\right] \nonumber \\
&\quad \times \left[
\begin{array} {cccc}
\overline {\cal T}_{1}^{*,I\leftarrow I} & \overline {\cal T}_{1}^{*,I\leftarrow Q} & \overline {\cal T}_{1}^{*,I\leftarrow U} & \overline {\cal T}_{1}^{*,I\leftarrow V} \\
\overline {\cal T}_{1}^{*,Q\leftarrow I} & \overline {\cal T}_{1}^{*,Q\leftarrow Q} & \overline {\cal T}_{1}^{*,Q\leftarrow U} & \overline {\cal T}_{1}^{*,Q\leftarrow V} \\
\overline {\cal T}_{1}^{*,U\leftarrow I} & \overline {\cal T}_{1}^{*,U\leftarrow Q} & \overline {\cal T}_{1}^{*,U\leftarrow U} & \overline {\cal T}_{1}^{*,U\leftarrow V} \\
\overline {\cal T}_{1}^{*,V\leftarrow I} & \overline {\cal T}_{1}^{*,V\leftarrow Q} & \overline {\cal T}_{1}^{*,V\leftarrow U} & \overline {\cal T}_{1}^{*,V\leftarrow V} \\
\end{array}
\right]\overline{\mathbf{X}}_{2}^{1}
\left[
\begin{array} {cccc}
\overline{\cal R}_{2}^{I\leftarrow I} & \overline{\cal R}_{2}^{I\leftarrow Q} & \overline{\cal R}_{2}^{I\leftarrow U} & \overline{\cal R}_{2}^{I\leftarrow V} \\
\overline{\cal R}_{2}^{Q\leftarrow I} & \overline{\cal R}_{2}^{Q\leftarrow Q} & \overline{\cal R}_{2}^{Q\leftarrow U} & \overline{\cal R}_{2}^{Q\leftarrow V} \\
\overline{\cal R}_{2}^{U\leftarrow I} & \overline{\cal R}_{2}^{U\leftarrow Q} & \overline{\cal R}_{2}^{U\leftarrow U} & \overline{\cal R}_{2}^{U\leftarrow V} \\
\overline{\cal R}_{2}^{V\leftarrow I} & \overline{\cal R}_{2}^{V\leftarrow Q} & \overline{\cal R}_{2}^{V\leftarrow U} & \overline{\cal R}_{2}^{V\leftarrow V} \\
\end{array}
\right] \nonumber \\
&\quad \times \left[
\begin{array} {cccc}
\overline {\cal T}_{1}^{I\leftarrow I} & \overline {\cal T}_{1}^{I\leftarrow Q} & \overline {\cal T}_{1}^{I\leftarrow U} & \overline {\cal T}_{1}^{I\leftarrow V} \\
\overline {\cal T}_{1}^{Q\leftarrow I} & \overline {\cal T}_{1}^{Q\leftarrow Q} & \overline {\cal T}_{1}^{Q\leftarrow U} & \overline {\cal T}_{1}^{Q\leftarrow V} \\
\overline {\cal T}_{1}^{U\leftarrow I} & \overline {\cal T}_{1}^{U\leftarrow Q} & \overline {\cal T}_{1}^{U\leftarrow U} & \overline {\cal T}_{1}^{U\leftarrow V} \\
\overline {\cal T}_{1}^{V\leftarrow I} & \overline {\cal T}_{1}^{V\leftarrow Q} & \overline {\cal T}_{1}^{V\leftarrow U} & \overline {\cal T}_{1}^{V\leftarrow V} \\
\end{array}
\right]
\end{align}
\end{subequations}
For a light beam incident from below, $\mathbf{R}^{*}_{1,2}(\mu_{0})$, $\mathbf{T}^{*}_{1,2}(\mu_{0})$,
$\overline{\cal R}^{*}_{1,2}$, and $\overline {\cal T}^{*}_{1,2}$ can be obtained in a similar manner.

In the downward path calculation, we apply it to multiple layers (from layer 1 to layer $k$).
\begin{subequations}\label{eq:12}
\begin{align}
& \left[
\begin{array} {cccc}
\mathbf{T}^{I}_{1,k}  \\
\mathbf{T}^{Q}_{1,k}  \\
\mathbf{T}^{U}_{1,k}  \\
\mathbf{T}^{V}_{1,k} \\
\end{array}
\right]=\left[
\begin{array} {cccc}
\mathbf{T}_{k}^{I}  \\
\mathbf{T}_{k}^{Q}  \\
\mathbf{T}_{k}^{U}  \\
\mathbf{T}_{k}^{V}  \\
\end{array}
\right]e^{-\frac{\tau_{1,k-1}}{\mu_{0}}} + \left[
\begin{array} {cccc}
\overline {\cal T}_{k}^{I\leftarrow I} & \overline {\cal T}_{k}^{I\leftarrow Q} & \overline {\cal T}_{k}^{I\leftarrow U} & \overline {\cal T}_{k}^{I\leftarrow V} \\
\overline {\cal T}_{k}^{Q\leftarrow I} & \overline {\cal T}_{k}^{Q\leftarrow Q} & \overline {\cal T}_{k}^{Q\leftarrow U} & \overline {\cal T}_{k}^{Q\leftarrow V} \\
\overline {\cal T}_{k}^{U\leftarrow I} & \overline {\cal T}_{k}^{U\leftarrow Q} & \overline {\cal T}_{k}^{U\leftarrow U} & \overline {\cal T}_{k}^{U\leftarrow V} \\
\overline {\cal T}_{k}^{V\leftarrow I} & \overline {\cal T}_{k}^{V\leftarrow Q} & \overline {\cal T}_{k}^{V\leftarrow U} & \overline {\cal T}_{k}^{V\leftarrow V} \\
\end{array}
\right] \left[
\begin{array} {cccc}
\mathbf{T}_{1,k-1}^{I}  \\
\mathbf{T}_{1,k-1}^{Q}  \\
\mathbf{T}_{1,k-1}^{U}  \\
\mathbf{T}_{1,k-1}^{V}  \\
\end{array}
\right] + \nonumber \\
&\quad \left[
\begin{array} {cccc}
\overline {\cal T}_{k}^{I\leftarrow I} & \overline {\cal T}_{k}^{I\leftarrow Q} & \overline {\cal T}_{k}^{I\leftarrow U} & \overline {\cal T}_{k}^{I\leftarrow V} \\
\overline {\cal T}_{k}^{Q\leftarrow I} & \overline {\cal T}_{k}^{Q\leftarrow Q} & \overline {\cal T}_{k}^{Q\leftarrow U} & \overline {\cal T}_{k}^{Q\leftarrow V} \\
\overline {\cal T}_{k}^{U\leftarrow I} & \overline {\cal T}_{k}^{U\leftarrow Q} & \overline {\cal T}_{k}^{U\leftarrow U} & \overline {\cal T}_{k}^{U\leftarrow V} \\
\overline {\cal T}_{k}^{V\leftarrow I} & \overline {\cal T}_{k}^{V\leftarrow Q} & \overline {\cal T}_{k}^{V\leftarrow U} & \overline {\cal T}_{k}^{V\leftarrow V} \\
\end{array}
\right]
\left[
\begin{array} {cccc}
\overline{\cal R}_{1,k-1}^{*,I\leftarrow I} & \overline{\cal R}_{1,k-1}^{*,I\leftarrow Q} & \overline{\cal R}_{1,k-1}^{*,I\leftarrow U} & \overline{\cal R}_{1,k-1}^{*,I\leftarrow V} \\
\overline{\cal R}_{1,k-1}^{*,Q\leftarrow I} & \overline{\cal R}_{1,k-1}^{*,Q\leftarrow Q} & \overline{\cal R}_{1,k-1}^{*,Q\leftarrow U} & \overline{\cal R}_{1,k-1}^{*,Q\leftarrow V} \\
\overline{\cal R}_{1,k-1}^{*,U\leftarrow I} & \overline{\cal R}_{1,k-1}^{*,U\leftarrow Q} & \overline{\cal R}_{1,k-1}^{*,U\leftarrow U} & \overline{\cal R}_{1,k-1}^{*,U\leftarrow V} \\
\overline{\cal R}_{1,k-1}^{*,V\leftarrow I} & \overline{\cal R}_{1,k-1}^{*,V\leftarrow Q} & \overline{\cal R}_{1,k-1}^{*,V\leftarrow U} & \overline{\cal R}_{1,k-1}^{*,V\leftarrow V} \\
\end{array}
\right] \overline{\mathbf{X}}_{k}^{1,k-1}\mathbf{Y}_{k}^{1.k-1} \\
&\left[
\begin{array} {cccc}
\overline{\cal R}_{1,k}^{*,I\leftarrow I} & \overline{\cal R}_{1,k}^{*,I\leftarrow Q} & \overline{\cal R}_{1,k}^{*,I\leftarrow U} & \overline{\cal R}_{1,k}^{*,I\leftarrow V} \\
\overline{\cal R}_{1,k}^{*,Q\leftarrow I} & \overline{\cal R}_{1,k}^{*,Q\leftarrow Q} & \overline{\cal R}_{1,k}^{*,Q\leftarrow U} & \overline{\cal R}_{1,k}^{*,Q\leftarrow V} \\
\overline{\cal R}_{1,k}^{*,U\leftarrow I} & \overline{\cal R}_{1,k}^{*,U\leftarrow Q} & \overline{\cal R}_{1,k}^{*,U\leftarrow U} & \overline{\cal R}_{1,k}^{*,U\leftarrow V} \\
\overline{\cal R}_{1,k}^{*,V\leftarrow I} & \overline{\cal R}_{1,k}^{*,V\leftarrow Q} & \overline{\cal R}_{1,k}^{*,V\leftarrow U} & \overline{\cal R}_{1,k}^{*,V\leftarrow V} \\
\end{array}
\right] =\left[
\begin{array} {cccc}
\overline{\cal R}_{k}^{*,I\leftarrow I} & \overline{\cal R}_{k}^{*,I\leftarrow Q} & \overline{\cal R}_{k}^{*,I\leftarrow U} & \overline{\cal R}_{k}^{*,I\leftarrow V} \\
\overline{\cal R}_{k}^{*,Q\leftarrow I} & \overline{\cal R}_{k}^{*,Q\leftarrow Q} & \overline{\cal R}_{k}^{*,Q\leftarrow U} & \overline{\cal R}_{k}^{*,Q\leftarrow V} \\
\overline{\cal R}_{k}^{*,U\leftarrow I} & \overline{\cal R}_{k}^{*,U\leftarrow Q} & \overline{\cal R}_{k}^{*,U\leftarrow U} & \overline{\cal R}_{k}^{*,U\leftarrow V} \\
\overline{\cal R}_{k}^{*,V\leftarrow I} & \overline{\cal R}_{k}^{*,V\leftarrow Q} & \overline{\cal R}_{k}^{*,V\leftarrow U} & \overline{\cal R}_{k}^{*,V\leftarrow V} \\
\end{array}
\right] \nonumber \\
&\quad +\left[
\begin{array} {cccc}
\overline {\cal T}_{k}^{I\leftarrow I} & \overline {\cal T}_{k}^{I\leftarrow Q} & \overline {\cal T}_{k}^{I\leftarrow U} & \overline {\cal T}_{k}^{I\leftarrow V} \\
\overline {\cal T}_{k}^{Q\leftarrow I} & \overline {\cal T}_{k}^{Q\leftarrow Q} & \overline {\cal T}_{k}^{Q\leftarrow U} & \overline {\cal T}_{k}^{Q\leftarrow V} \\
\overline {\cal T}_{k}^{U\leftarrow I} & \overline {\cal T}_{k}^{U\leftarrow Q} & \overline {\cal T}_{k}^{U\leftarrow U} & \overline {\cal T}_{k}^{U\leftarrow V} \\
\overline {\cal T}_{k}^{V\leftarrow I} & \overline {\cal T}_{k}^{V\leftarrow Q} & \overline {\cal T}_{k}^{V\leftarrow U} & \overline {\cal T}_{k}^{V\leftarrow V} \\
\end{array}
\right]\overline{\mathbf{X}}_{k}^{*1,k-1}
\left[
\begin{array} {cccc}
\overline{\cal R}_{1,k-1}^{*,I\leftarrow I} & \overline{\cal R}_{1,k-1}^{*,I\leftarrow Q} & \overline{\cal R}_{1,k-1}^{*,I\leftarrow U} & \overline{\cal R}_{1,k-1}^{*,I\leftarrow V} \\
\overline{\cal R}_{1,k-1}^{*,Q\leftarrow I} & \overline{\cal R}_{1,k-1}^{*,Q\leftarrow Q} & \overline{\cal R}_{1,k-1}^{*,Q\leftarrow U} & \overline{\cal R}_{1,k-1}^{*,Q\leftarrow V} \\
\overline{\cal R}_{1,k-1}^{*,U\leftarrow I} & \overline{\cal R}_{1,k-1}^{*,U\leftarrow Q} & \overline{\cal R}_{1,k-1}^{*,U\leftarrow U} & \overline{\cal R}_{1,k-1}^{*,U\leftarrow V} \\
\overline{\cal R}_{1,k-1}^{*,V\leftarrow I} & \overline{\cal R}_{1,k-1}^{*,V\leftarrow Q} & \overline{\cal R}_{1,k-1}^{*,V\leftarrow U} & \overline{\cal R}_{1,k-1}^{*,V\leftarrow V} \\
\end{array}
\right] \nonumber \\
&\quad \times \left[
\begin{array} {cccc}
\overline {\cal T}_{k}^{*,I\leftarrow I} & \overline {\cal T}_{k}^{*,I\leftarrow Q} & \overline {\cal T}_{k}^{*,I\leftarrow U} & \overline {\cal T}_{k}^{*,I\leftarrow V} \\
\overline {\cal T}_{k}^{*,Q\leftarrow I} & \overline {\cal T}_{k}^{*,Q\leftarrow Q} & \overline {\cal T}_{k}^{*,Q\leftarrow U} & \overline {\cal T}_{k}^{*,Q\leftarrow V} \\
\overline {\cal T}_{k}^{*,U\leftarrow I} & \overline {\cal T}_{k}^{*,U\leftarrow Q} & \overline {\cal T}_{k}^{*,U\leftarrow U} & \overline {\cal T}_{k}^{*,U\leftarrow V} \\
\overline {\cal T}_{k}^{*,V\leftarrow I} & \overline {\cal T}_{k}^{*,V\leftarrow Q} & \overline {\cal T}_{k}^{*,V\leftarrow U} & \overline {\cal T}_{k}^{*,V\leftarrow V} \\
\end{array}
\right]
\end{align}
\end{subequations}
where $\overline{\mathbf{X}}_{i}^{*j}$ is provided in Appendix B.

Similarly, in an upward path calculation, we apply the multi-layer from the surface to layer $k+1$:
\begin{subequations}\label{eq:13}
\begin{align}
&\left[
\begin{array} {cccc}
\mathbf{R}^{I}_{k,N}  \\
\mathbf{R}^{Q}_{k,N}  \\
\mathbf{R}^{U}_{k,N}  \\
\mathbf{R}^{V}_{k,N} \\
\end{array}
\right]=\left[
\begin{array} {cccc}
\mathbf{R}_{k}^{I}  \\
\mathbf{R}_{k}^{Q}  \\
\mathbf{R}_{k}^{U}  \\
\mathbf{R}_{k}^{V}  \\
\end{array}
\right] + \left[
\begin{array} {cccc}
\overline {\cal T}_{k}^{*,I\leftarrow I} & \overline {\cal T}_{k}^{*,I\leftarrow Q} & \overline {\cal T}_{k}^{*,I\leftarrow U} & \overline {\cal T}_{k}^{*,I\leftarrow V} \\
\overline {\cal T}_{k}^{*,Q\leftarrow I} & \overline {\cal T}_{k}^{*,Q\leftarrow Q} & \overline {\cal T}_{k}^{*,Q\leftarrow U} & \overline {\cal T}_{k}^{*,Q\leftarrow V} \\
\overline {\cal T}_{k}^{*,U\leftarrow I} & \overline {\cal T}_{k}^{*,U\leftarrow Q} & \overline {\cal T}_{k}^{*,U\leftarrow U} & \overline {\cal T}_{k}^{*,U\leftarrow V} \\
\overline {\cal T}_{k}^{*,V\leftarrow I} & \overline {\cal T}_{k}^{*,V\leftarrow Q} & \overline {\cal T}_{k}^{*,V\leftarrow U} & \overline {\cal T}_{k}^{*,V\leftarrow V} \\
\end{array}
\right]\overline{\mathbf{X}}_{k}^{k+1,N}\mathbf{Y}_{k}^{k+1,N} \\
& \left[
\begin{array} {cccc}
\overline{\cal R}_{k,N}^{I\leftarrow I} & \overline{\cal R}_{k,N}^{I\leftarrow Q} & \overline{\cal R}_{k,N}^{I\leftarrow U} & \overline{\cal R}_{k,N}^{I\leftarrow V} \\
\overline{\cal R}_{k,N}^{Q\leftarrow I} & \overline{\cal R}_{k,N}^{Q\leftarrow Q} & \overline{\cal R}_{k,N}^{Q\leftarrow U} & \overline{\cal R}_{k,N}^{Q\leftarrow V} \\
\overline{\cal R}_{k,N}^{U\leftarrow I} & \overline{\cal R}_{k,N}^{U\leftarrow Q} & \overline{\cal R}_{k,N}^{U\leftarrow U} & \overline{\cal R}_{k,N}^{U\leftarrow V} \\
\overline{\cal R}_{k,N}^{V\leftarrow I} & \overline{\cal R}_{k,N}^{V\leftarrow Q} & \overline{\cal R}_{k,N}^{V\leftarrow U} & \overline{\cal R}_{k,N}^{V\leftarrow V} \\
\end{array}
\right] =\left[
\begin{array} {cccc}
\overline{\cal R}_{k}^{I\leftarrow I} & \overline{\cal R}_{k}^{I\leftarrow Q} & \overline{\cal R}_{k}^{I\leftarrow U} & \overline{\cal R}_{k}^{I\leftarrow V} \\
\overline{\cal R}_{k}^{Q\leftarrow I} & \overline{\cal R}_{k}^{Q\leftarrow Q} & \overline{\cal R}_{k}^{Q\leftarrow U} & \overline{\cal R}_{k}^{Q\leftarrow V} \\
\overline{\cal R}_{k}^{U\leftarrow I} & \overline{\cal R}_{k}^{U\leftarrow Q} & \overline{\cal R}_{k}^{U\leftarrow U} & \overline{\cal R}_{k}^{U\leftarrow V} \\
\overline{\cal R}_{k}^{V\leftarrow I} & \overline{\cal R}_{k}^{V\leftarrow Q} & \overline{\cal R}_{k}^{V\leftarrow U} & \overline{\cal R}_{k}^{V\leftarrow V} \\
\end{array}
\right] \nonumber \\
& \quad + \left[
\begin{array} {cccc}
\overline {\cal T}_{k}^{*,I\leftarrow I} & \overline {\cal T}_{k}^{*,I\leftarrow Q} & \overline {\cal T}_{k}^{*,I\leftarrow U} & \overline {\cal T}_{k}^{*,I\leftarrow V} \\
\overline {\cal T}_{k}^{*,Q\leftarrow I} & \overline {\cal T}_{k}^{*,Q\leftarrow Q} & \overline {\cal T}_{k}^{*,Q\leftarrow U} & \overline {\cal T}_{k}^{*,Q\leftarrow V} \\
\overline {\cal T}_{k}^{*,U\leftarrow I} & \overline {\cal T}_{k}^{*,U\leftarrow Q} & \overline {\cal T}_{k}^{*,U\leftarrow U} & \overline {\cal T}_{k}^{*,U\leftarrow V} \\
\overline {\cal T}_{k}^{*,V\leftarrow I} & \overline {\cal T}_{k}^{*,V\leftarrow Q} & \overline {\cal T}_{k}^{*,V\leftarrow U} & \overline {\cal T}_{k}^{*,V\leftarrow V} \\
\end{array}
\right]\overline{\mathbf{X}}_{k}^{k+1,N}
\left[
\begin{array} {cccc}
\overline{\cal R}_{k+1,N}^{I\leftarrow I} & \overline{\cal R}_{k+1,N}^{I\leftarrow Q} & \overline{\cal R}_{k+1,N}^{I\leftarrow U} & \overline{\cal R}_{k+1,N}^{I\leftarrow V} \\
\overline{\cal R}_{k+1,N}^{Q\leftarrow I} & \overline{\cal R}_{k+1,N}^{Q\leftarrow Q} & \overline{\cal R}_{k+1,N}^{Q\leftarrow U} & \overline{\cal R}_{k+1,N}^{Q\leftarrow V} \\
\overline{\cal R}_{k+1,N}^{U\leftarrow I} & \overline{\cal R}_{k+1,N}^{U\leftarrow Q} & \overline{\cal R}_{k+1,N}^{U\leftarrow U} & \overline{\cal R}_{k+1,N}^{U\leftarrow V} \\
\overline{\cal R}_{k+1,N}^{V\leftarrow I} & \overline{\cal R}_{k+1,N}^{V\leftarrow Q} & \overline{\cal R}_{k+1,N}^{V\leftarrow U} & \overline{\cal R}_{k+1,N}^{V\leftarrow V} \\
\end{array}
\right] \nonumber \\
&\quad \times
\left[
\begin{array} {cccc}
\overline {\cal T}_{k}^{I\leftarrow I} & \overline {\cal T}_{k}^{I\leftarrow Q} & \overline {\cal T}_{k}^{I\leftarrow U} & \overline {\cal T}_{k}^{I\leftarrow V} \\
\overline {\cal T}_{k}^{Q\leftarrow I} & \overline {\cal T}_{k}^{Q\leftarrow Q} & \overline {\cal T}_{k}^{Q\leftarrow U} & \overline {\cal T}_{k}^{Q\leftarrow V} \\
\overline {\cal T}_{k}^{U\leftarrow I} & \overline {\cal T}_{k}^{U\leftarrow Q} & \overline {\cal T}_{k}^{U\leftarrow U} & \overline {\cal T}_{k}^{U\leftarrow V} \\
\overline {\cal T}_{k}^{V\leftarrow I} & \overline {\cal T}_{k}^{V\leftarrow Q} & \overline {\cal T}_{k}^{V\leftarrow U} & \overline {\cal T}_{k}^{V\leftarrow V} \\
\end{array}
\right]
\end{align}
\end{subequations}
where $\mathbf{R}_{N}=
\left[\begin{array} {cccc}
 \textbf{R}_{N}^{I} \\
 \textbf{R}_{N}^{Q} \\
 \textbf{R}_{N}^{U} \\
 \textbf{R}_{N}^{V} \\
\end{array}
\right]$
is the surface albedo and $\overline{\cal R}_{N}=
\left[
\begin{array} {cccc}
\overline{\boldsymbol{\mathcal{R}}}_{N}^{I\leftarrow I}  & \overline{\boldsymbol{\mathcal{R}}}_{N}^{I\leftarrow Q} &
\overline{\boldsymbol{\mathcal{R}}}_{N}^{I\leftarrow U}  & \overline{\boldsymbol{\mathcal{R}}}_{N}^{I\leftarrow V} \\
\overline{\boldsymbol{\mathcal{R}}}_{N}^{Q\leftarrow I}  & \overline{\boldsymbol{\mathcal{R}}}_{N}^{Q\leftarrow Q} &
\overline{\boldsymbol{\mathcal{R}}}_{N}^{Q\leftarrow U}  & \overline{\boldsymbol{\mathcal{R}}}_{N}^{Q\leftarrow V} \\
\overline{\boldsymbol{\mathcal{R}}}_{N}^{U\leftarrow I}  & \overline{\boldsymbol{\mathcal{R}}}_{N}^{U\leftarrow Q} &
\overline{\boldsymbol{\mathcal{R}}}_{N}^{U\leftarrow U}  & \overline{\boldsymbol{\mathcal{R}}}_{N}^{U\leftarrow V} \\
\overline{\boldsymbol{\mathcal{R}}}_{N}^{V\leftarrow I}  & \overline{\boldsymbol{\mathcal{R}}}_{N}^{V\leftarrow Q} &
\overline{\boldsymbol{\mathcal{R}}}_{N}^{V\leftarrow U}  & \overline{\boldsymbol{\mathcal{R}}}_{N}^{V\leftarrow V} \\
\end{array}
\right]$ is the reflection property of the surface.

Thus, the internal Stokes parameters at level $k+1$ can be determined as follows:
\begin{subequations}\label{eq:14}
\begin{align}
&\left[
\begin{array} {cccc}
\mathbf{A}^{I}_{k+1}  \\
\mathbf{A}^{Q}_{k+1}  \\
\mathbf{A}^{U}_{k+1}  \\
\mathbf{A}^{V}_{k+1} \\
\end{array}
\right]=\overline{\mathbf{X}}_{1,k}^{k+1,N}\mathbf{Y}_{1,k}^{k+1,N}, \\
&\left[
\begin{array} {cccc}
\mathbf{D}^{I}_{k+1}  \\
\mathbf{D}^{Q}_{k+1}  \\
\mathbf{D}^{U}_{k+1}  \\
\mathbf{D}^{V}_{k+1} \\
\end{array}
\right]=\left[
\begin{array} {cccc}
\mathbf{T}_{1,k}^{I}  \\
\mathbf{T}_{1,k}^{Q}  \\
\mathbf{T}_{1,k}^{U}  \\
\mathbf{T}_{1,k}^{V}  \\
\end{array}
\right]+\left[
\begin{array} {cccc}
\overline{\cal R}_{1,k}^{*,I\leftarrow I} & \overline{\cal R}_{1,k}^{*,I\leftarrow Q} & \overline{\cal R}_{1,k}^{*,I\leftarrow U} & \overline{\cal R}_{1,k}^{*,I\leftarrow V} \\
\overline{\cal R}_{1,k}^{*,Q\leftarrow I} & \overline{\cal R}_{1,k}^{*,Q\leftarrow Q} & \overline{\cal R}_{1,k}^{*,Q\leftarrow U} & \overline{\cal R}_{1,k}^{*,Q\leftarrow V} \\
\overline{\cal R}_{1,k}^{*,U\leftarrow I} & \overline{\cal R}_{1,k}^{*,U\leftarrow Q} & \overline{\cal R}_{1,k}^{*,U\leftarrow U} & \overline{\cal R}_{1,k}^{*,U\leftarrow V} \\
\overline{\cal R}_{1,k}^{*,V\leftarrow I} & \overline{\cal R}_{1,k}^{*,V\leftarrow Q} & \overline{\cal R}_{1,k}^{*,V\leftarrow U} & \overline{\cal R}_{1,k}^{*,V\leftarrow V} \\
\end{array}
\right]
\overline{\mathbf{X}}_{1,k}^{k+1,N}\mathbf{Y}_{1,k}^{k+1,N}.
\end{align}
\end{subequations}

An adjusted absorption and scattering atmosphere may be considered to incorporate the forward peak contribution into multiple scattering. The fraction of the scattered energy residing in the forward peak $f$ is separated from the phase function. This is the $\delta$-M adjustment \citep{Spurr2006} used in the proposed model.

The intensity at an arbitrary zenith angle for satellite applications can be obtained by replacing $\mu_i$ in Eq.(4) and the satellite zenith angle $\mu_{sat}$.
The entire calculation process is the same for $I(\tau,\pm\mu_{sat})$, $Q(\tau,\pm\mu_{sat})$, $U(\tau,\pm\mu_{sat})$, and $V(\tau,\pm\mu_{sat})$. We call this new method the polarized discrete ordinate adding approximation (POLDDA) for vector solar radiative transfer.

In the PolRadtran/RT3 model, which is a code based on the adding-doubling method, the adding method is always used together with the doubling method because the multiple scattering process is not considered in the single-layer solution and the calculation time increases with an increase in the optical thickness. However, in POLDDA, no doubling process is required and the adding method can be used directly after a single-layer solution (considering multiple scattering processes). Therefore, in theory, the calculation time does not change with the variation in the optical depth. This point is demonstrated in Section 4.3.

\section{Comparison Results}
In this section, the calculation accuracy and efficiency of POLDDA are evaluated by comparing with the Monte Carlo model (MYSTIC) \citep{Emde2015} and PolRadtran/RT3, where the MYSTIC model is considered the benchmark. Test cases for both single-layer and multi-layer atmospheres are included. POLDDA and RT3 were executed with 16 streams (half-sphere) for all Rayleigh cases, and 32 streams/64 streams (half-sphere) for the water cloud case.
\subsection{Single-layer test cases}
In the first set of cases, one layer including molecules of atmospheric constituents with different depolarization factors and no surface reflection was tested. Another case exists, which includes Lambertian surface reflection.
\subsubsection{Case 1 - Rayleigh scattering with different depolarization factor}
Case 1 (including cases 1-1, 1-2, 1-3) was used to verify the correct treatment of the anisotropy of the molecules through the Rayleigh depolarization factor. It contains one layer of non-absorbing molecules (single scattering albedo $\omega=1$) and an optical thickness of 0.5. Surface reflection was not considered (the surface albedo was 0). Stokes vectors were calculated for various depolarization factors and sun positions. The definition of the viewing zenith angle is with respect to the downward normal instead of the upward normal, so viewing zenith angles are 0$^{\circ}$ -80$^{\circ}$ (down-looking) at the bottom and 100$^{\circ}$-180$^{\circ}$ (up-looking) at the top with 5$^{\circ}$ increments. The results at viewing zenith angle near 90$^{\circ}$ (direction near horizontal) are not shown. The viewing azimuthal angle definition is clockwise and ranged from 0$^{\circ}$ to 360$^{\circ}$ (5$^{\circ}$ increments).

Test case 1-1 had the simplest setup with a zero Rayleigh depolarization factor. The Stokes vectors at the bottom and TOA were calculated for solar zenith angles $\theta_{0}=$0$^{\circ}$ and $\phi_{0}=$65$^{\circ}$, respectively. The left plots shown in Figure 2 present the results calculated using POLDDA. The absolute errors and relative differences between POLDDA and RT3 compared to the MYSTIC model are shown. The $U$-component is zero because the sun was located at the zenith. From Fig. 2, the relative differences by POLDDA are smaller than 0.1\% for the $I$-component, which is almost the same as that of RT3 in both the upward and downward directions. The related bias are almost less than 2\% for the $Q$-component at the top and bottom of the atmosphere using both POLDDA and RT3. In one point, the abnormally large related bias is shown at viewing azimuth angles of 180$^{\circ}$ (top) and 0$^{\circ}$ (bottom). It is because the values of $Q$-component are unsuitable as a denominator for calculating relative error if they are near zero. From the absolute error figures, a large error does not exist, which is a good illustration.  When we removed abnormally large relative differences from Fig.2, the relative difference plot was consistent with the absolute difference plot (as shown in Fig. 2).
The same phenomenon also appeared in subsequent cases, but we did not remove the deviation again.
The relative root mean square errors (RMSE) for all the test cases are listed in Table 1. We calculated the RMSE for the $I$-component between MYSTIC and POLDDA or RT3 using RMSE=$\frac{\sqrt{\sum_{i=1}^{N}(I_{MYSTIC}^{i}-I_{o}^{i})^{2}}}{\sqrt{\sum_{i=1}^{N}(I_{MYSTIC}^{i})^{2}}}$ (where $I_{o}$ refers to the other models) for the radiation field including all down-and up-looking directions. The RMSEs of POLDDA and RT3 are 0.0177\% for the $I$-component and 0.0226\% for the $Q$-component, which also demonstrates the accuracy of POLDDA in this case.

Test cases 1-2 and 1-3 consider two cases in which the depolarization factor is not zero; it is 0.03 in case 1-2 and 0.1 in case 1-3. The solar positions ($\theta_{0}=$30$^{\circ}$, $\phi_{0}$=65$^{\circ}$) are the same in cases 1-2 and 1-3. The $U$-component is nonzero because the solar zenith angle is not zero. Figures 3 and 4 present the Stokes vector results at the TOA and bottom, respectively. We show the relative bias, and as shown in Fig. 3, the relative bias for the $I$-component of POLDDA is between -0.05\% and 0.05\%. The largest bias of the $Q$-component reached a high of 2\%, ranging from 120$^{\circ}$-150$^{\circ}$ and near 170$^{\circ}$ and 285$^{\circ}$ viewing azimuth angles, for both POLDDA and RT3. The $U$-component calculated by POLDDA and RT3 did not show a significant difference, except near viewing azimuth angles of 0$^{\circ}$ and 180$^{\circ}$ owing to the $U$-component near zero. The large bias of the $I$, $Q$, and $U$ components yielded by POLDDA at the bottom (Fig. 4) is almost the same as that of RT3. For Case 1-2, the RMSEs of POLDDA were close to those of RT3 (Table 1).

In Case 1-3, although the value of the depolarization factor increased, the bias of POLDDA (Figures 5 and 6) also differed slightly from that of RT3. The RMSE of POLDDA are 0.0166\%, 0.025\%, and 0.0229\% (Table 1) for the $I$-, $Q$-, and $U$-components, respectively, which are similar to those of RT3.

\subsubsection{Case 2 - Rayleigh atmosphere with Lambertian surface}
Surface reflection was considered in Case 2 with a Lambertian surface albedo of 0.3, as well as for a one-layer medium of 0.1 without absorption. The optical depth of the layer was set as 0.1. The solar position was $\theta_{0}=$50$^{\circ}$, with $\phi_{0}$=0$^{\circ}$, and the Rayleigh depolarization factor is the same as that in Case 1-3.
The Stokes vectors ($I$-, $Q$-, $U$-, and $V$-components) were calculated for viewing zenith angles of 0-80$^{\circ}$ (100-180$^{\circ}$) at the bottom (top) with 5$^{\circ}$ increments, and viewing azimuth angles of 0-180$^{\circ}$ with 5$^{\circ}$ increments.

Figure 7 shows the results of Case 2. The relative bias of POLDDA is less than 0.05\% for the $I$-component. A relative bias greater than 1\% occurs when viewing azimuth angles are approximately 170\ degrees at the top of the layer for the $Q$-component and approximately 30\ degrees at the bottom. For the $U$-component, over 5\% relative bias is observed in the viewing azimuth angles ranging from 0 to 10$^{\circ}$ at the bottom and 100$^{\circ}$-110$^{\circ}$ at the top of the layer. Significant differences were not found in the $I$-, $Q$-, and $U$-components calculated using POLDDA and RT3. The RMSEs of POLDDA and RT3 were also similar to those of nonzero Lambertian surface reflection.

\subsection{Test cases for realistic atmosphere with multi-layer}
In this set of cases, the coupling between POLDDA model layers was examined using a multi-layer in a plane-parallel atmosphere. We used the addition method to handle multi-layer connections.
Unlike RT3, the doubling process for POLDDA is not required; only the addition process is used with single-layer reflection and transmission functions.

The U.S. standard atmosphere \citep{McClatchey1972} was used for the multi-layer cases. The model atmosphere comprised 30 layers from 0 km to 30 km in an altitude with a thickness of 1 km.
The surface albedo was set to 0 for Cases 3 and 4 and 0.2 for Case 5. The Rayleigh scattering depolarization factor is 0.03. The sun?¡¥s position is $\theta_{0}=$60$^{\circ}$ and $\phi_{0}=$0$^{\circ}$. For Cases 3 and 4, the radiance field was calculated at the surface (0 km) and TOA (30 km) for viewing zenith angles from 0$^{\circ}$ to 85$^{\circ}$ (up-looking), and 95$^{\circ}$ to 180$^{\circ}$ (down-looking) in 5$^{\circ}$ increments. The viewing azimuth angle ranged from 0$^{\circ}$ to 180$^{\circ}$ in  5$^{\circ}$ increments. For Case 5, the radiance field was calculated at the surface (0 km) and TOA (30 km) for viewing zenith angles from 0$^{\circ}$ to 90$^{\circ}$ (up-looking) and 90$^{\circ}$ to 180$^{\circ}$ (down-looking) at a 90$^{\circ}$ viewing azimuth angle.

\subsubsection{Case 3 - Only Rayleigh scattering with standard atmosphere}
The radiance field was calculated as 450 nm. Only Rayleigh scattering was considered.
The profiles of the scattering optical thicknesses are depicted in Figure 8(a) and can be downloaded from the International Polarized Radiative Transfer (IPRT) website.

Figure 9 shows the $I$-, $Q$-, and $U$-components at the output altitudes for a multi-layer atmosphere. For the one-layer cases, good agreement between the three models was observed for pure Rayleigh scattering with zero depolarization factor. In the inhomogeneous multi-layer, the three models are in good agreement, except near 0$^{\circ}$-10$^{\circ}$ at the bottom and 100$^{\circ}$-110$^{\circ}$ viewing azimuth angles at TOA for the $U$-component. The bias of POLDDA is  less than 0.2\% for the $I$-component and 0.5\% for the $Q$-component. The RMSE of POLDDA for the $I$-, $Q$-, and $U$-components are only 0.017\%, 0.0266\%, and 0.0198\%, respectively, against MYSTIC (Table 1).

\subsubsection{Case 4 - Rayleigh scattering and absorption with standard atmosphere}
In this case, it was determined whether the absorption effect was appropriately considered.
The radiance field was calculated at a wavelength of 325 nm, where the ozone absorption is strong.
The profiles of scattering optical thicknesses and absorption optical thicknesses are illustrated in Figures 8 (b-c), which can also be downloaded from the IPRT website.

Figure 10 shows the $I$-, $Q$-, and $U$-component results at the top and on the surface. Similar to Case 3, little difference exists as calculated by POLDDA and RT3 for $I$-, $Q$-, and $U$-components. The difference between RMSE of POLDDA and RT3 decreases compared to the values in Case 4 owing to the addition of gas absorption, particularly for the $I$-component.

\subsubsection{Case 5 - Standard atmosphere with cloud layer}
In this case, a cloud layer was added between 2 and 3 km. The radiance was measured at 800 nm. The cloud optical thickness is 5, which is larger than the atmosphere that only includes air molecules. The profiles of the Rayleigh scattering optical thicknesses with no molecular absorption are shown in Figure 8(d). The single-scattering albedo for the cloud layer is 0.999979 and the cloud phase function was calculated using Mie scattering. POLDDA and RT3 were executed with 32 and 64 streams (half-sphere), respectively.

Figures 11 and 12 show the $I$-, $Q$-, $U$-, and $V$-component results for RT3 (64 streams), and the relative differences between POLDDA and RT3 at the top and surface, respectively. The relative differences between POLDDA and RT3 are less than 0.01\%, 0.025\%, 0.05\%, and 0.05\% for the $I$-, $Q$-, $U$-, and $V$-components at TOA (Fig. 11), respectively, and are notably close to zero as $|cos\theta|$ decreases. At the surface (Fig. 12), the relative difference between POLDDA and RT3 is close to zero when $|cos\theta|>0.1$. Thus, in the case of water clouds, the results of POLDDA and RT3 agree.

\subsection{Computational efficiency}
Although the single-layer solution of POLDDA is complicated by the consideration of multiple scattering compared to RT3, POLDDA also saves time in the operation of the doubling process. To illustrate the efficiency of POLDDA, we compare the efficiency of POLDDA with that of RT3. Figure 13(a) shows the computational times calculated using the POLDDA and RT3 models with increasing optical depth. A two-layer atmosphere with only molecular scattering is considered, and the optical depth of each layer is in the range from 0.001 to 100. The half-stream was 16, as calculated using POLDDA and RT3. Admittedly, the RT3 model has better computational efficiency when the optical depth is less than 0.0001 because it ignores multiple scattering effects when solving the equation of a single-layer atmosphere. However, the time run by RT3 increases with optical depth. The computational time increases by more than four times when the optical depth reaches 100. Unlike RT3, the computing time of POLDDA does not increase with increasing optical depth of each layer. The operation-time advantage of POLDDA is gradually apparent when the optical depth exceeds 0.0005. Following this trend, POLDDA saves more computational time than RT3 in aerosols or clouded atmospheres (optically thicker than 1). The computational efficiencies of POLDDA and RT3 with different half streams are also considered (Fig. 13b). The layer optical depth was set as 1. As shown in Fig. 13(b), the computational time increases with an increasing number of streams for both POLDDA and RT3, whereas the computational time for RT3 is longer than that for POLDDA with an increasing number of streams.

\section{Conclusions and discussion}
In this study, the polarized adding method of discrete ordinate approximation (POLDDA) was developed for ultraviolet-visible and near-infrared spectra. The single-layer polarized radiative transfer equation and inhomogeneous multilayer connection are solved using the discrete ordinate method and adding method, respectively. From the accuracy evaluation results in a multiple-layer standard atmosphere with Rayleigh scattering and a cloud layer, POLDDA proved to conform to the results of PolRadtran/RT3. The RMSE values of the Stokes vectors between POLDDA and RT3 against MYSTIC were found to be similar, which further confirms the good accuracy of POLDDA. POLDDA also has a high computational efficiency, particularly for an atmosphere with an optical depth of over 0.0005 compared to RT3. Moreover, the computation time of POLDDA does not increase with the optical depth of each layer.

\section*{Appendix A}
\setcounter{equation}{0}
\renewcommand{\theequation}{A.\arabic{equation}}

From Eq.(1), we expand the phase matrix and $\mathbf{L}$ into the Fourier cosine and sine series,
\begin{subequations} \label{eq:A1}
\begin{eqnarray}
\left[
\begin{array} {c}
I(\tau,\mu,\varphi) \\
Q(\tau,\mu,\varphi) \\
U(\tau,\mu,\varphi) \\
V(\tau,\mu,\varphi) \\
\end{array}
\right]&=&\sum^{2M-1}_{m=0}
\left[
\begin{array} {c}
I^{m}(\tau,\mu)\cdot cos(\varphi-\varphi_{0}) \\
Q^{m}(\tau,\mu)\cdot cos(\varphi-\varphi_{0}) \\
U^{m}(\tau,\mu)\cdot sin(\varphi-\varphi_{0}) \\
V^{m}(\tau,\mu)\cdot sin(\varphi-\varphi_{0}) \\
\end{array}
\right] \\
\mathbf{Z}&=&
\left[
\begin{array} {cccc}
z_{11} & z_{12} & z_{13} & z_{14} \\
z_{21} & z_{22} & z_{23} & z_{24} \\
z_{31} & z_{32} & z_{33} & z_{34} \\
z_{41} & z_{42} & z_{43} & z_{44}
\end{array}
\right]
\end{eqnarray}
\end{subequations}
where
$z_{xy}(\mu,\varphi,\mu',\varphi')=\sum^{2M-1}_{m=0}\left\{
\begin{array} {lr}
z^{m}_{xy}(\mu,\varphi,\mu',\varphi')\cdot cosm(\varphi-\varphi')  \\
z^{m}_{xy}(\mu,\varphi,\mu',\varphi')\cdot sinm(\varphi-\varphi')  \\
\end{array}
\right.$
Thus, the radiation transfer equation can be split into 2M equations.

The discrete ordinate method is used to solve the equation. To achieve this, a Gaussian quadrature was used to handle the integration in Eq.(1) as
\begin{equation} \label{eq:A2}
\int_{-1} ^1 \mathbf{Z}^{m}(\mu,\mu')\mathbf{L}^{m}(\tau,\mu')d\mu'=
\sum^{N}_{j=-N,j\neq0}a_{j}\mathbf{Z}^{m}(\mu ,\mu_{j})\mathbf{L}^{m}(\tau ,\mu_{j})
\end{equation}
where $N$ denotes the number of half-streams, and $\mu_j = -\mu_{-j}$ and $a_j = a_{-j}$ $(j=1,2,...,N)$ denote the quadrature angles and weights, respectively.

Upon substituting Eqs.(A1)-(A2) into Eq.(1) and neglecting the superscript $m$, Eq.(1) can be written as follows:
\begin{equation} \label{eq:A3}
\begin{split}
\mu \frac{d\mathbf{L}(\tau,\mu_{i})}{d\tau} =& \mathbf{L}(\tau,\mu_{i}) - \frac{\omega}{4}(1+\delta_{0,m}) \sum^{N}_{j=-N,j\neq0}a_{j}\mathbf{Z}(\mu_{i},\mu_{j})\mathbf{L}(\tau ,\mu_{j}) \\
&-\frac{\omega}{4\pi}\mathbf{Z}(\mu_{i},-\mu_{0})\mathbf{F_{0}}e^{-\tau/\mu_{0}}.
\end{split}
\end{equation}
where $N$ indicates the number of half-streams, and $\mu_j = -\mu_{-j}$ and $a_j = a_{-j}$ $(j=1,2,...,N)$ denote the quadrature angles and weights, respectively. They can be written in matrix form as Eq.(2).

\section*{Appendix B}
In Eq.(7),
\begin{equation} \nonumber
\begin{array} {lcrc}
\textbf{A}^{I}=
\left[
\begin{array} {rc}
R^{I}(\tau_{1},\mu_{1}),\cdots,R^{I}(\tau_{1},\mu_{N})
\end{array}
\right]_{N\times1}^{\mathbb{T}},
\textbf{A}^{Q}=
\left[
\begin{array} {rc}
R^{Q}(\tau_{1},\mu_{1}), \cdots,R^{Q}(\tau_{1},\mu_{N})
\end{array}
\right]_{N\times1}^{\mathbb{T}},\\
\textbf{A}^{U}=
\left[
\begin{array} {rc}
R^{U}(\tau_{1},\mu_{1}),\cdots,R^{U}(\tau_{1},\mu_{N})
\end{array}
\right]_{N\times1}^{\mathbb{T}},
\textbf{A}^{V}=
\left[
\begin{array} {rc}
R^{V}(\tau_{1},\mu_{1}),\cdots,R^{V}(\tau_{1},\mu_{N})
\end{array}
\right]_{N\times1}^{\mathbb{T}}, \\
\textbf{D}^{I}=\left[
\begin{array} {rc}
T^{I}(\tau_{1},-\mu_{1}),\cdots,  T^{I}(\tau_{1},-\mu_{N})
\end{array}\right]_{N\times1}^{\mathbb{T}},
\textbf{D}^{Q}=
\left[
\begin{array} {rc}
T^{Q}(\tau_{1},-\mu_{1}), \cdots,T^{Q}(\tau_{1},-\mu_{N})
\end{array}\right]_{N\times1}^\mathbb{T}, \\
\textbf{D}^{U}=\left[
\begin{array} {rc}
T^{U}(\tau_{1},-\mu_{1}),\cdots,  T^{U}(\tau_{1},-\mu_{N})
\end{array}\right]_{N\times1}^{\mathbb{T}},
\textbf{D}^{V}=
\left[
\begin{array} {rc}
T^{V}(\tau_{1},-\mu_{1}), \cdots,T^{V}(\tau_{1},-\mu_{N})
\end{array}\right]_{N\times1}^\mathbb{T}
\end{array}
\end{equation}
and others matrix are
$\mathbf{R}_{1}=[\mathbf{R}_{1}^{I},\mathbf{R}_{1}^{Q},\mathbf{R}_{1}^{U},\mathbf{R}_{1}^{V}]^\mathbb{T}$, $\mathbf{T}_{1}=[\mathbf{T}_{1}^{I},\mathbf{T}_{1}^{Q},\mathbf{T}_{1}^{U},\mathbf{T}_{1}^{V}]^\mathbb{T}$,\\
$\mathbf{R}_{2}=[\mathbf{R}_{2}^{I},\mathbf{R}_{2}^{Q},\mathbf{R}_{2}^{U},\mathbf{R}_{2}^{V}]^\mathbb{T}$,
$\mathbf{T}_{2}=[\mathbf{T}_{2}^{I},\mathbf{T}_{2}^{Q},\mathbf{T}_{2}^{U},\mathbf{T}_{2}^{V}]^\mathbb{T}$,\\
\begin{equation} \nonumber
\overline{\mathbf{R}}_{2}=\left[
\begin{array} {cccc}
\overline {\cal R}_{2}^{I\leftarrow I} & \overline {\cal R}_{2}^{I\leftarrow Q} & \overline {\cal R}_{2}^{I\leftarrow U} & \overline {\cal R}_{2}^{I\leftarrow V} \\
\overline {\cal R}_{2}^{Q\leftarrow I} & \overline {\cal R}_{2}^{Q\leftarrow Q} & \overline {\cal R}_{2}^{Q\leftarrow U} & \overline {\cal R}_{2}^{Q\leftarrow V} \\
\overline {\cal R}_{2}^{U\leftarrow I} & \overline {\cal R}_{2}^{U\leftarrow Q} & \overline {\cal R}_{2}^{U\leftarrow U} & \overline {\cal R}_{2}^{U\leftarrow V} \\
\overline {\cal R}_{2}^{V\leftarrow I} & \overline {\cal R}_{2}^{V\leftarrow Q} & \overline {\cal R}_{2}^{V\leftarrow U} & \overline {\cal R}_{2}^{V\leftarrow V} \\
\end{array}
\right]
\end{equation} \\
\begin{equation} \nonumber
\overline{\mathbf{R}}^{*}_{1}=\left[
\begin{array} {cccc}
\overline {\cal R}_{1}^{*,I\leftarrow I} & \overline {\cal R}_{1}^{*,I\leftarrow Q} & \overline {\cal R}_{1}^{*,I\leftarrow U} & \overline {\cal R}_{1}^{*,I\leftarrow V} \\
\overline {\cal R}_{1}^{*,Q\leftarrow I} & \overline {\cal R}_{1}^{*,Q\leftarrow Q} & \overline {\cal R}_{1}^{*,Q\leftarrow U} & \overline {\cal R}_{1}^{*,Q\leftarrow V} \\
\overline {\cal R}_{1}^{*,U\leftarrow I} & \overline {\cal R}_{1}^{*,U\leftarrow Q} & \overline {\cal R}_{1}^{*,U\leftarrow U} & \overline {\cal R}_{1}^{*,U\leftarrow V} \\
\overline {\cal R}_{1}^{*,V\leftarrow I} & \overline {\cal R}_{1}^{*,V\leftarrow Q} & \overline {\cal R}_{1}^{*,V\leftarrow U} & \overline {\cal R}_{1}^{*,V\leftarrow V} \\
\end{array}
\right]
\end{equation} \\
\begin{equation} \nonumber
\overline{\mathbf{T}}^{*}_{1}=\left[
\begin{array} {cccc}
\overline {\cal T}_{1}^{*,I\leftarrow I} & \overline {\cal T}_{1}^{*,I\leftarrow Q} & \overline {\cal T}_{1}^{*,I\leftarrow U} & \overline {\cal T}_{1}^{*,I\leftarrow V} \\
\overline {\cal T}_{1}^{*,Q\leftarrow I} & \overline {\cal T}_{1}^{*,Q\leftarrow Q} & \overline {\cal T}_{1}^{*,Q\leftarrow U} & \overline {\cal T}_{1}^{*,Q\leftarrow V} \\
\overline {\cal T}_{1}^{*,U\leftarrow I} & \overline {\cal T}_{1}^{*,U\leftarrow Q} & \overline {\cal T}_{1}^{*,U\leftarrow U} & \overline {\cal T}_{1}^{*,U\leftarrow V} \\
\overline {\cal T}_{1}^{*,V\leftarrow I} & \overline {\cal T}_{1}^{*,V\leftarrow Q} & \overline {\cal T}_{1}^{*,V\leftarrow U} & \overline {\cal T}_{1}^{*,V\leftarrow V} \\
\end{array}
\right]
\end{equation} \\
\begin{equation} \nonumber
\overline{\mathbf{T}}_{2}=\left[
\begin{array} {cccc}
\overline {\cal T}_{2}^{I\leftarrow I} & \overline {\cal T}_{2}^{I\leftarrow Q} & \overline {\cal T}_{2}^{I\leftarrow U} & \overline {\cal T}_{2}^{I\leftarrow V} \\
\overline {\cal T}_{2}^{Q\leftarrow I} & \overline {\cal T}_{2}^{Q\leftarrow Q} & \overline {\cal T}_{2}^{Q\leftarrow U} & \overline {\cal T}_{2}^{Q\leftarrow V} \\
\overline {\cal T}_{2}^{U\leftarrow I} & \overline {\cal T}_{2}^{U\leftarrow Q} & \overline {\cal T}_{2}^{U\leftarrow U} & \overline {\cal T}_{2}^{U\leftarrow V} \\
\overline {\cal T}_{2}^{V\leftarrow I} & \overline {\cal T}_{2}^{V\leftarrow Q} & \overline {\cal T}_{2}^{V\leftarrow U} & \overline {\cal T}_{2}^{V\leftarrow V} \\
\end{array}
\right]
\end{equation}

In Eq.(9),
\begin{equation} \nonumber
\begin{array} {lcrc}
\overline{\mathbf{X}}_{i}^{j}= \{\mathbf{E} - \left[
\begin{array} {cccc}
\overline{\cal R}_{i}^{I\leftarrow I} & \overline{\cal R}_{i}^{I\leftarrow Q} & \overline{\cal R}_{i}^{I\leftarrow U} & \overline{\cal R}_{i}^{I\leftarrow V} \\
\overline{\cal R}_{i}^{Q\leftarrow I} & \overline{\cal R}_{i}^{Q\leftarrow Q} & \overline{\cal R}_{i}^{Q\leftarrow U} & \overline{\cal R}_{i}^{Q\leftarrow V} \\
\overline{\cal R}_{i}^{U\leftarrow I} & \overline{\cal R}_{i}^{U\leftarrow Q} & \overline{\cal R}_{i}^{U\leftarrow U} & \overline{\cal R}_{i}^{U\leftarrow V} \\
\overline{\cal R}_{i}^{V\leftarrow I} & \overline{\cal R}_{i}^{V\leftarrow Q} & \overline{\cal R}_{i}^{V\leftarrow U} & \overline{\cal R}_{i}^{V\leftarrow V} \\
\end{array}
\right]
\left[
\begin{array} {cccc}
\overline{\cal R}_{j}^{*,I\leftarrow I} & \overline{\cal R}_{j}^{*,I\leftarrow Q} & \overline{\cal R}_{j}^{*,I\leftarrow U} & \overline{\cal R}_{j}^{*,I\leftarrow V} \\
\overline{\cal R}_{j}^{*,Q\leftarrow I} & \overline{\cal R}_{j}^{*,Q\leftarrow Q} & \overline{\cal R}_{j}^{*,Q\leftarrow U} & \overline{\cal R}_{j}^{*,Q\leftarrow V} \\
\overline{\cal R}_{j}^{*,U\leftarrow I} & \overline{\cal R}_{j}^{*,U\leftarrow Q} & \overline{\cal R}_{j}^{*,U\leftarrow U} & \overline{\cal R}_{j}^{*,U\leftarrow V} \\
\overline{\cal R}_{j}^{*,V\leftarrow I} & \overline{\cal R}_{j}^{*,V\leftarrow Q} & \overline{\cal R}_{j}^{*,V\leftarrow U} & \overline{\cal R}_{j}^{*,V\leftarrow V} \\
\end{array}
\right]\}
\\
\mathbf{Y}_{i}^{j}=\{\left[
\begin{array} {cccc}
\mathbf{R}_{i}^{I}  \\
\mathbf{R}_{i}^{Q}  \\
\mathbf{R}_{i}^{U}  \\
\mathbf{R}_{i}^{V}  \\
\end{array}
\right]e^{-\frac{\tau_{j}}{\mu_{0}}} + \left[
\begin{array} {cccc}
\overline{\cal R}_{i}^{I\leftarrow I} & \overline{\cal R}_{i}^{I\leftarrow Q} & \overline{\cal R}_{i}^{I\leftarrow U} & \overline{\cal R}_{i}^{I\leftarrow V} \\
\overline{\cal R}_{i}^{Q\leftarrow I} & \overline{\cal R}_{i}^{Q\leftarrow Q} & \overline{\cal R}_{i}^{Q\leftarrow U} & \overline{\cal R}_{i}^{Q\leftarrow V} \\
\overline{\cal R}_{i}^{U\leftarrow I} & \overline{\cal R}_{i}^{U\leftarrow Q} & \overline{\cal R}_{i}^{U\leftarrow U} & \overline{\cal R}_{i}^{U\leftarrow V} \\
\overline{\cal R}_{i}^{V\leftarrow I} & \overline{\cal R}_{i}^{V\leftarrow Q} & \overline{\cal R}_{i}^{V\leftarrow U} & \overline{\cal R}_{i}^{V\leftarrow V} \\
\end{array}
\right]
\left[
\begin{array} {cccc}
\mathbf{T}_{j}^{I}  \\
\mathbf{T}_{j}^{Q}  \\
\mathbf{T}_{j}^{U}  \\
\mathbf{T}_{j}^{V}  \\
\end{array}
\right]\}
\end{array}
\end{equation}
and $\mathbf{E}$ is a 4N$\times$4N identity matrix.

In Eq.(12),
\begin{equation} \nonumber
\begin{array} {lcrc}
\overline{\mathbf{X}}_{i}^{*j}= \{\mathbf{E} - \left[
\begin{array} {cccc}
\overline{\cal R}_{i}^{*,I\leftarrow I} & \overline{\cal R}_{i}^{*,I\leftarrow Q} & \overline{\cal R}_{i}^{*,I\leftarrow U} & \overline{\cal R}_{i}^{*,I\leftarrow V} \\
\overline{\cal R}_{i}^{*,Q\leftarrow I} & \overline{\cal R}_{i}^{*,Q\leftarrow Q} & \overline{\cal R}_{i}^{*,Q\leftarrow U} & \overline{\cal R}_{i}^{*,Q\leftarrow V} \\
\overline{\cal R}_{i}^{*,U\leftarrow I} & \overline{\cal R}_{i}^{*,U\leftarrow Q} & \overline{\cal R}_{i}^{*,U\leftarrow U} & \overline{\cal R}_{i}^{*,U\leftarrow V} \\
\overline{\cal R}_{i}^{*,V\leftarrow I} & \overline{\cal R}_{i}^{*,V\leftarrow Q} & \overline{\cal R}_{i}^{*,V\leftarrow U} & \overline{\cal R}_{i}^{*,V\leftarrow V} \\
\end{array}
\right]
\left[
\begin{array} {cccc}
\overline{\cal R}_{j}^{I\leftarrow I} & \overline{\cal R}_{j}^{I\leftarrow Q} & \overline{\cal R}_{j}^{I\leftarrow U} & \overline{\cal R}_{j}^{I\leftarrow V} \\
\overline{\cal R}_{j}^{Q\leftarrow I} & \overline{\cal R}_{j}^{Q\leftarrow Q} & \overline{\cal R}_{j}^{Q\leftarrow U} & \overline{\cal R}_{j}^{Q\leftarrow V} \\
\overline{\cal R}_{j}^{U\leftarrow I} & \overline{\cal R}_{j}^{U\leftarrow Q} & \overline{\cal R}_{j}^{U\leftarrow U} & \overline{\cal R}_{j}^{U\leftarrow V} \\
\overline{\cal R}_{j}^{V\leftarrow I} & \overline{\cal R}_{j}^{V\leftarrow Q} & \overline{\cal R}_{j}^{V\leftarrow U} & \overline{\cal R}_{j}^{V\leftarrow V} \\
\end{array}
\right]\}
\end{array}
\end{equation}

\section*{Acknowledgments}
The code of POLDDA is available from the corresponding author upon reasonable request.
This study was supported by the National Natural Science Foundation of China (42222506 and 42105081) and China Postdoctoral Science Foundation (2023M730618).

\bibliography{references}
\newpage

\begin{table}[t]
\caption{ Relative root mean square errors in percent between MYSTIC and POLDDA, RT3 for all cases in this study.}
\label{t1}
\begin{center}
\begin{tabular}{ccccccccccc}
  \hline
  Case     &   & POLDDA     & RT3 \\
  \hline
  Case1-1  & I & 0.01773589\% & 0.01773660\% \\
           & Q & 0.02265512\% & 0.02264818\% \\
  \hline
  Case1-2  & I & 0.01649993\% & 0.01650125\% \\
           & Q & 0.02561625\% & 0.02561332\% \\
           & U & 0.02343666\% & 0.02343717\% \\
  \hline
  Case1-3  & I & 0.01660634\% & 0.01660778\% \\
           & Q & 0.02499878\% & 0.02499902\% \\
           & U & 0.02290941\% & 0.02290969\% \\
  \hline
  Case2  & I & 0.008561691\% & 0.008559853\% \\
         & Q & 0.03571449\% & 0.03571167\% \\
         & U & 0.02874306\% & 0.02874456\% \\
  \hline
  Case3  & I & 0.01723639\% & 0.01723684\% \\
         & Q & 0.02661624\% & 0.02660797\% \\
         & U & 0.01979122\% & 0.01979095\% \\
  \hline
  Case4  & I & 0.01260397\% & 0.01260261\% \\
         & Q & 0.02472782\% & 0.02472788\% \\
         & U & 0.02089377\% & 0.02090551\% \\
\hline
\end{tabular}
\end{center}
\end{table}

\begin{figure}\label{t2}
  \centerline{\includegraphics[width=36pc,angle=0]{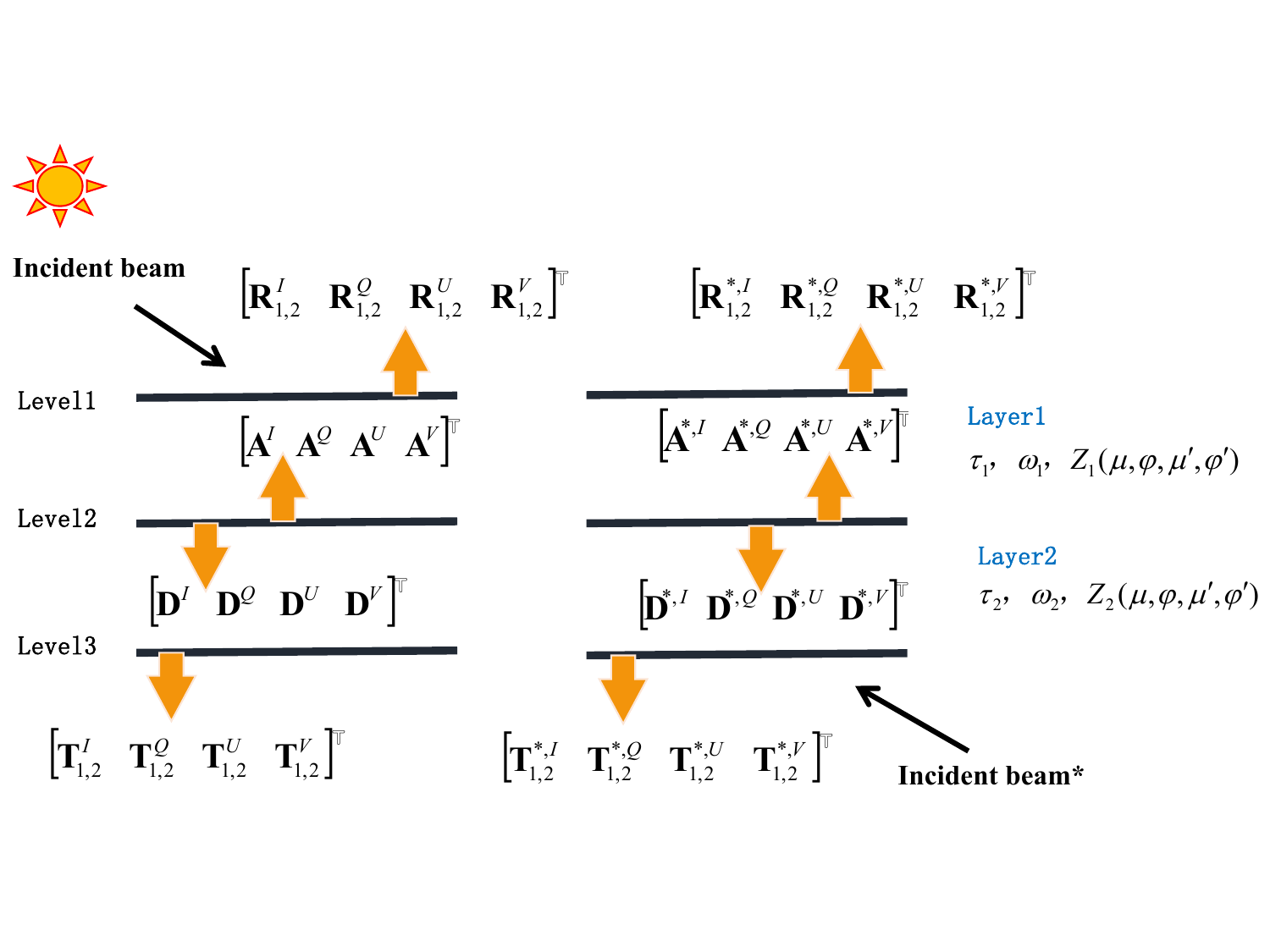}}
  \caption{Schematic diagram of the principles of invariance in vector solar radiative transfer.}
\end{figure}

\begin{figure}\label{t3}
  \centerline{\includegraphics[width=36pc,angle=0]{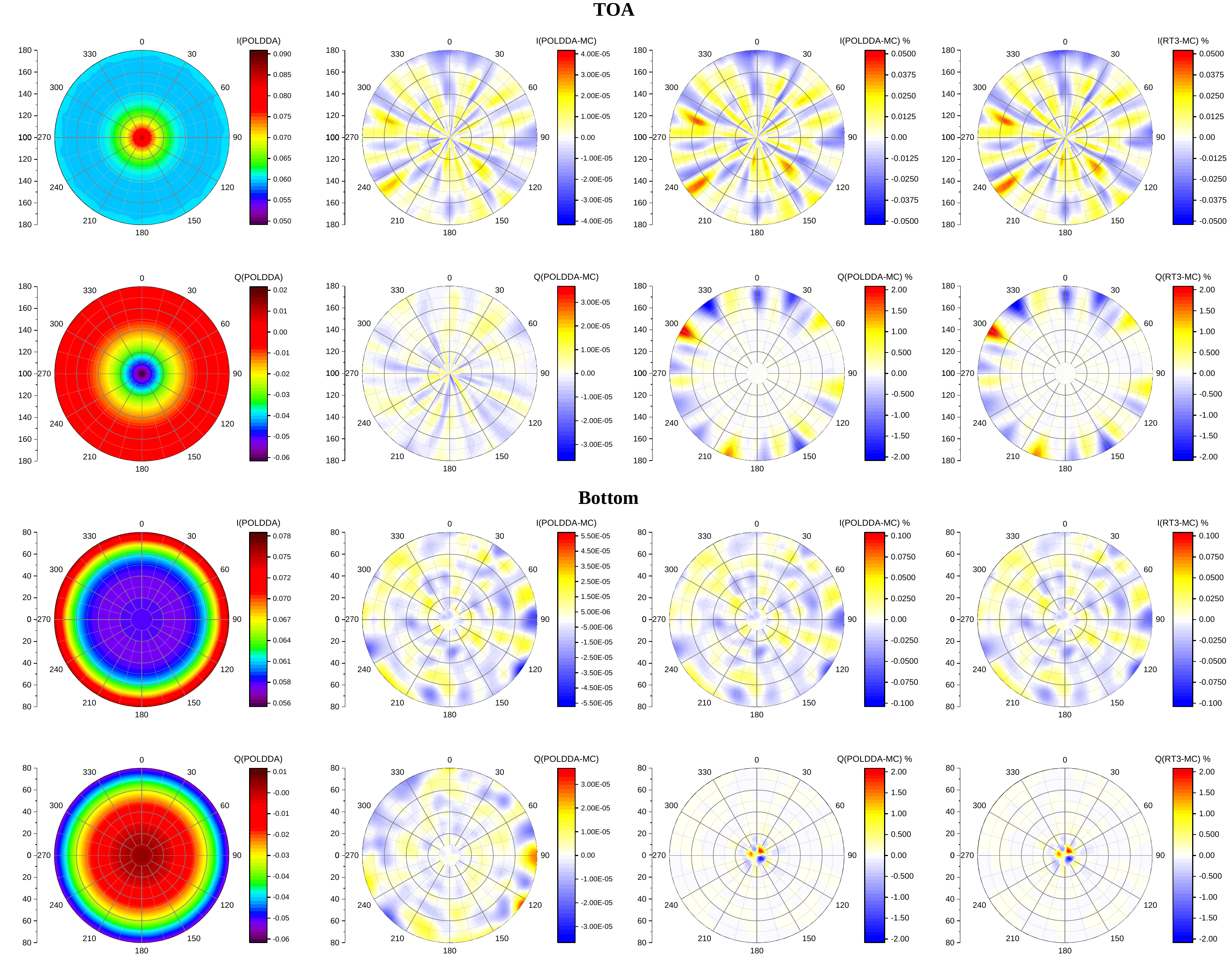}}
  \caption{Case 1-1: Rayleigh scattering, and depol is zero, (left column) the results for the $I$-component and $Q$-component at the top and bottom of the layer based on POLDDA (16S), units: $Wm^{-2}\mu m^{-1}sr^{-1}$; (second column to the left) the absolute differences of POLDDA (16S) against MYSTIC, units: $Wm^{-2}\mu m^{-1}sr^{-1}$; (third column to the left) the relative differences of POLDDA (16S) against MYSTIC and (right column) the relative differences of RT3 (16S) against MYSTIC, units: \%.}
\end{figure}

\begin{figure}\label{t4}
  \centerline{\includegraphics[width=40pc,angle=0]{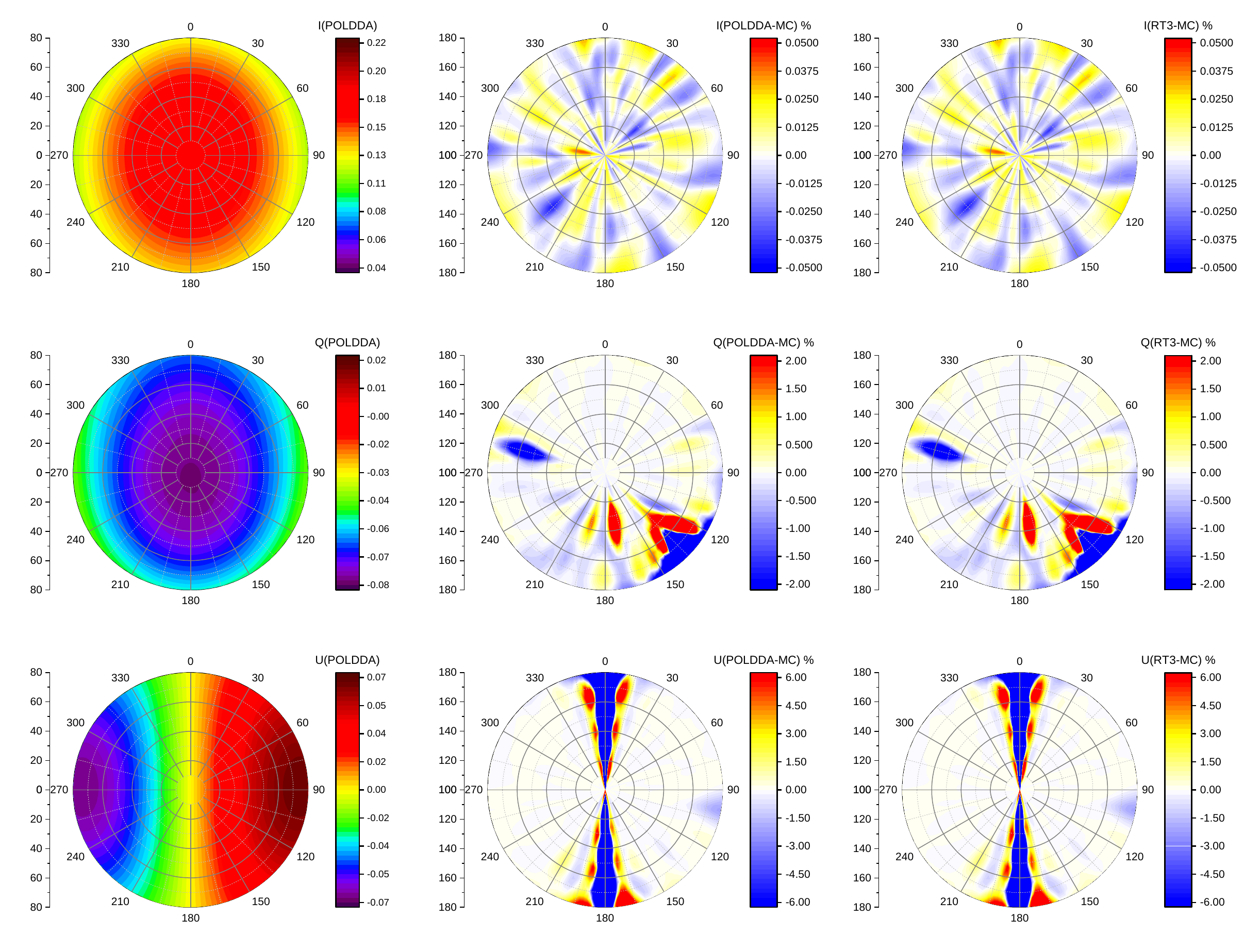}}
  \caption{Case 1-2: Rayleigh scattering, and depol is 0.03, (left column) the results for the $I$-component, $Q$-component, and $U$-component at the top of the layer based on POLDDA (16S), units: $Wm^{-2}\mu m^{-1}sr^{-1}$; (middle column) the relative differences of POLDDA (16S) against MYSTIC and (right column) the relative differences of RT3 (16S) against MYSTIC, units: \%.}
\end{figure}

\begin{figure}\label{t5}
  \centerline{\includegraphics[width=40pc,angle=0]{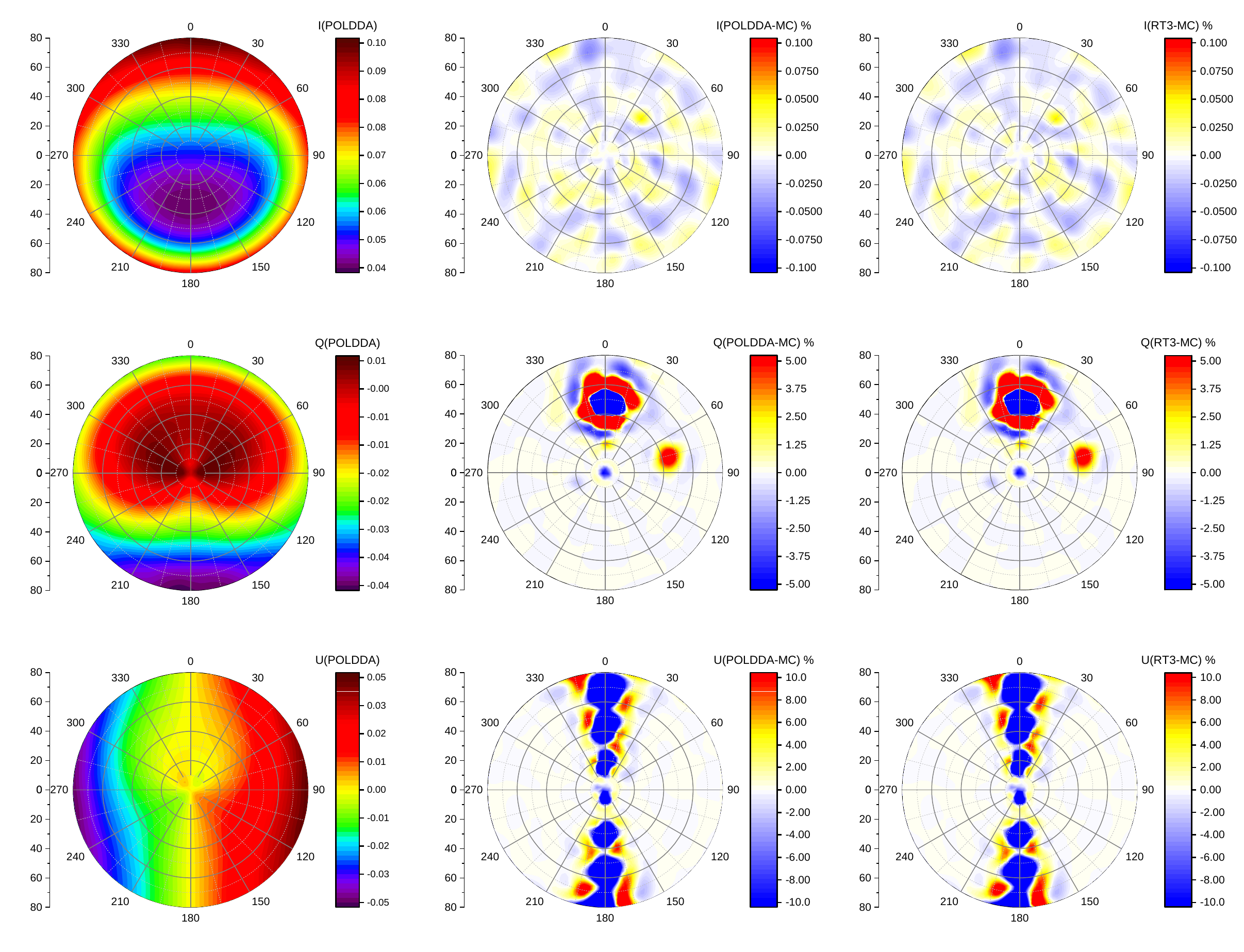}}
  \caption{Case 1-2: the same as Fig. 3, except for the bottom.}
\end{figure}

\begin{figure}\label{t6}
  \centerline{\includegraphics[width=40pc,angle=0]{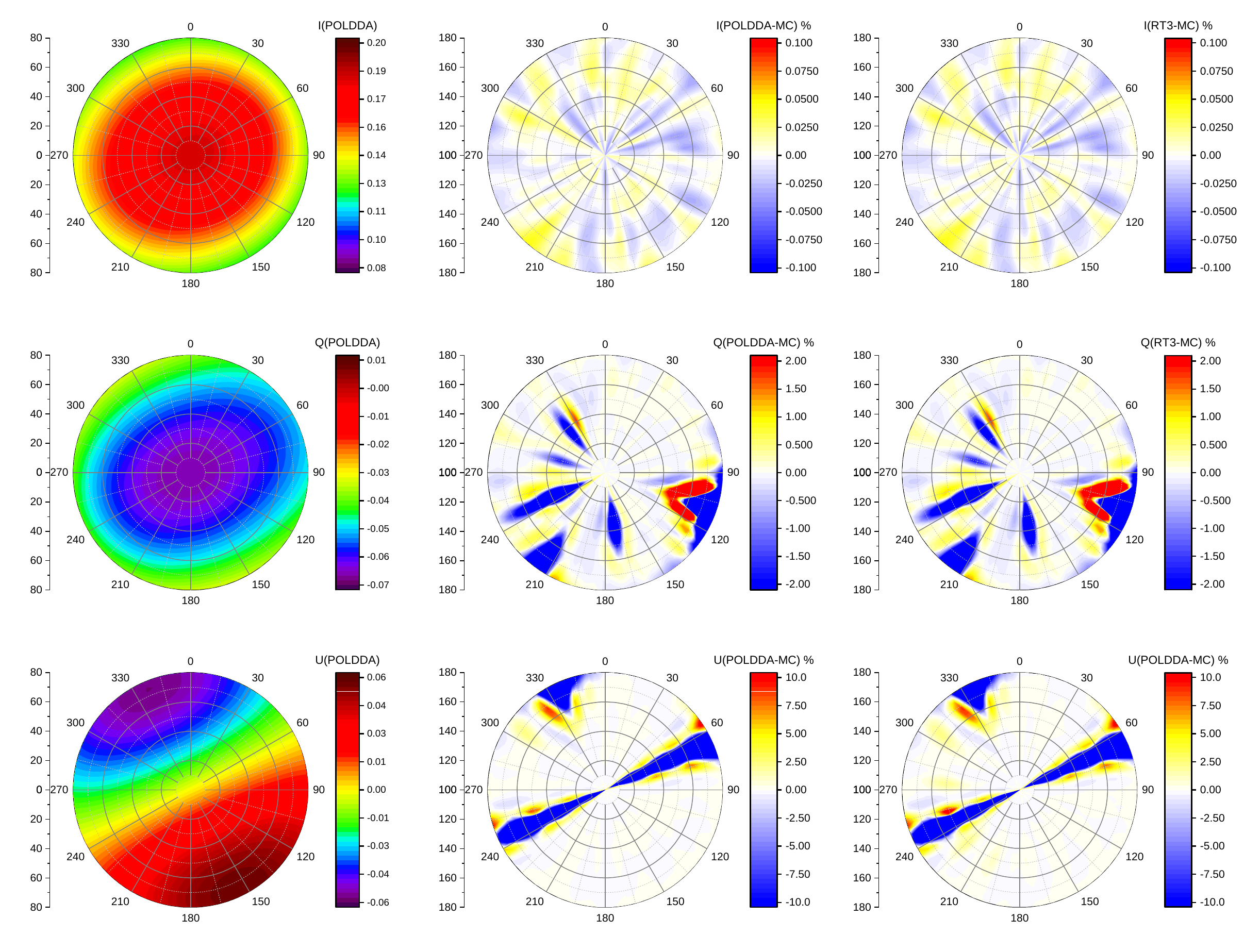}}
  \caption{Case 1-3: Rayleigh scattering, and depol is 0.1, (left column) the results for the $I$-component, $Q$-component, and $U$-component at the top of the layer based on POLDDA (16S), units: $Wm^{-2}\mu m^{-1}sr^{-1}$; (middle column) the relative differences of POLDDA (16S) against MYSTIC and (right column) the relative differences of RT3 (16S) against MYSTIC, units: \%.}
\end{figure}

\begin{figure}\label{t7}
  \centerline{\includegraphics[width=40pc,angle=0]{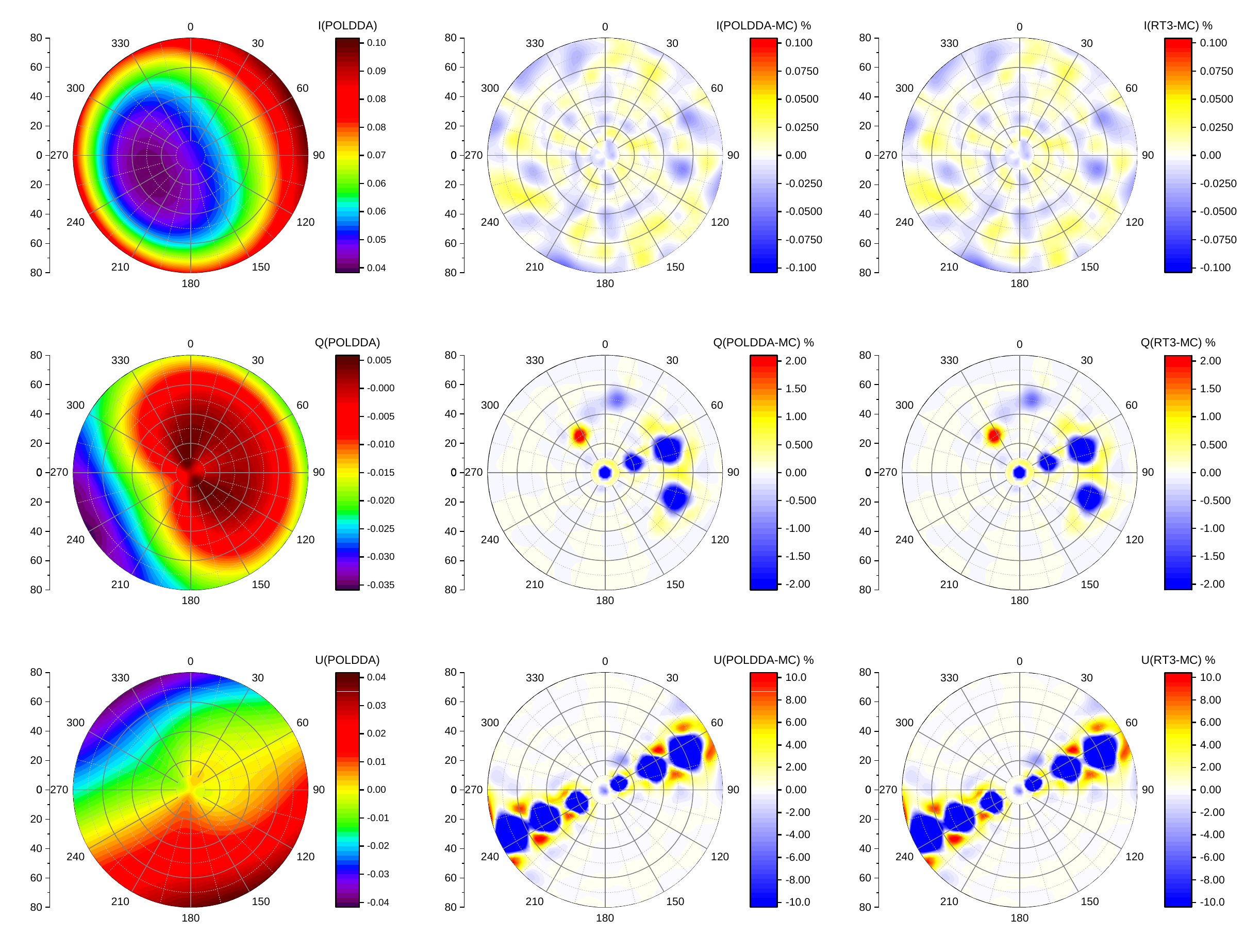}}
  \caption{Case 1-3: the same as Fig. 5, except for the bottom.}
\end{figure}

\begin{figure}\label{t8}
  \centerline{\includegraphics[width=40pc,angle=0]{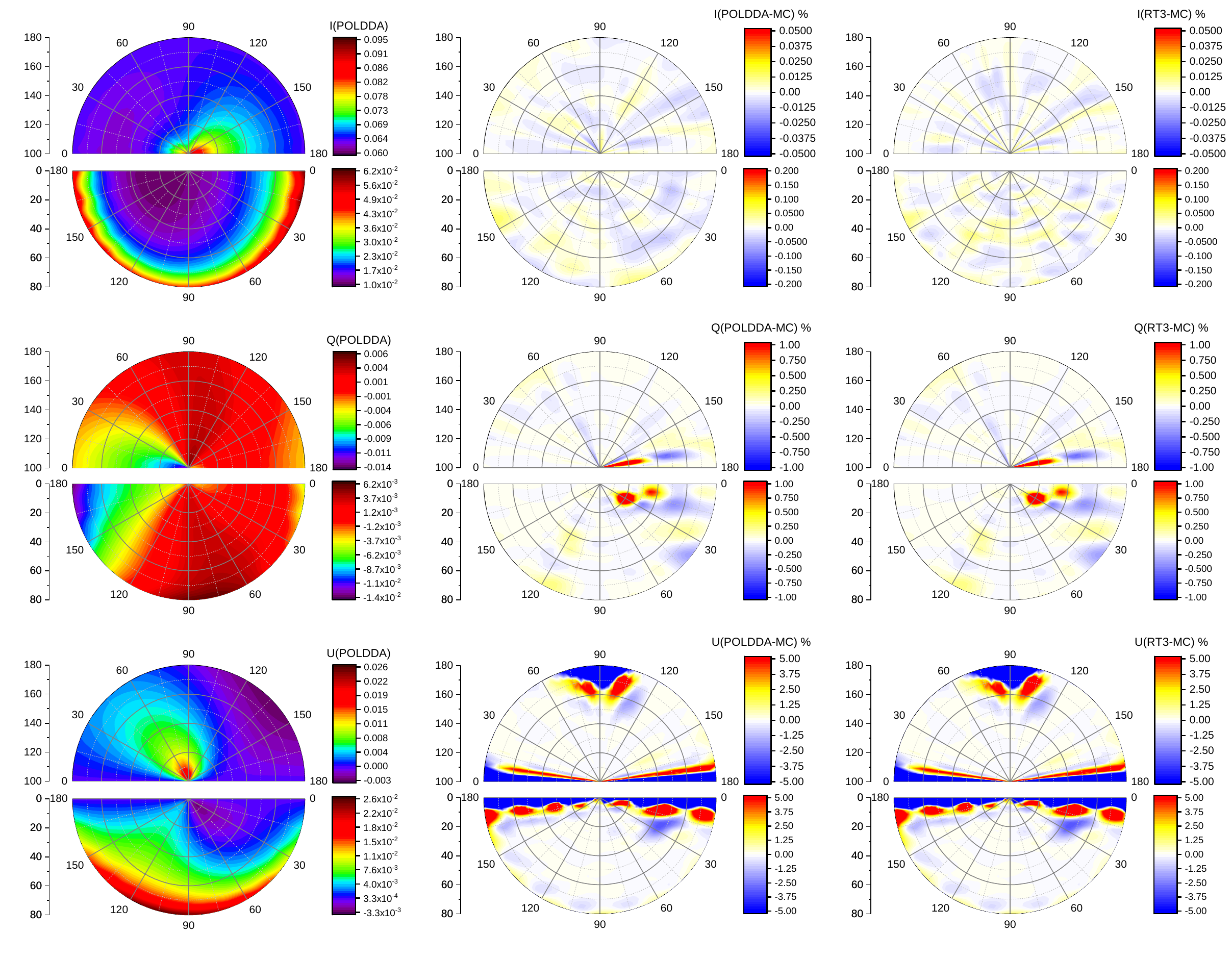}}
  \caption{Case 2: Rayleigh atmosphere with Lambertian surface, (left column) the results for the $I$-component, $Q$-component, and $U$-component at the top of the layer based on POLDDA (16S), units: $Wm^{-2}\mu m^{-1}sr^{-1}$; (middle column) the relative differences of POLDDA (16S) against MYSTIC, and (right column) the relative differences of RT3 (16S) against MYSTIC, units: \%.}
\end{figure}

\begin{figure}\label{t9}
  \centerline{\includegraphics[width=36pc,angle=0]{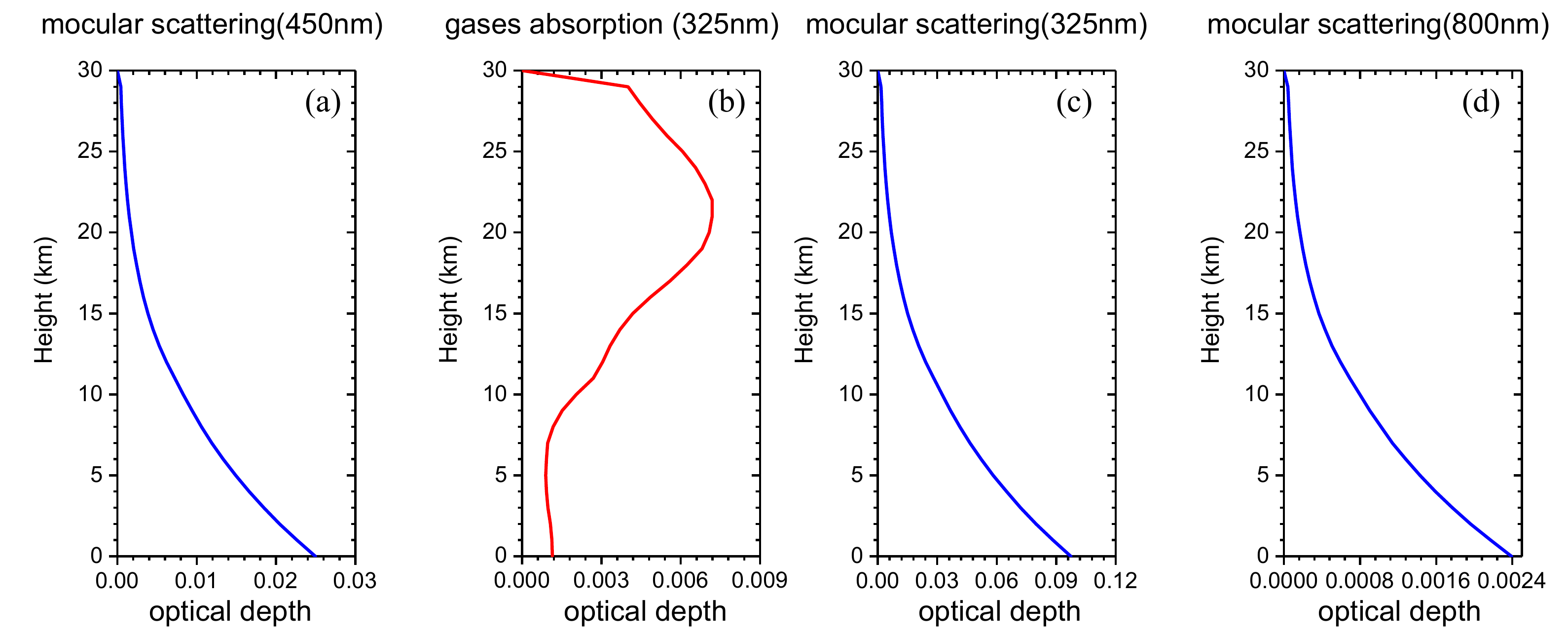}}
  \caption{Optical depth profiles for multi-layer test cases. The left plot shows the molecular scattering optical depth used in Case 3 (450 nm). The middle
  and the right plots show the gases absorption optical depth and molecular scattering optical depth in Case 4 (325 nm).}
\end{figure}

\begin{figure}\label{t10}
  \centerline{\includegraphics[width=40pc,angle=0]{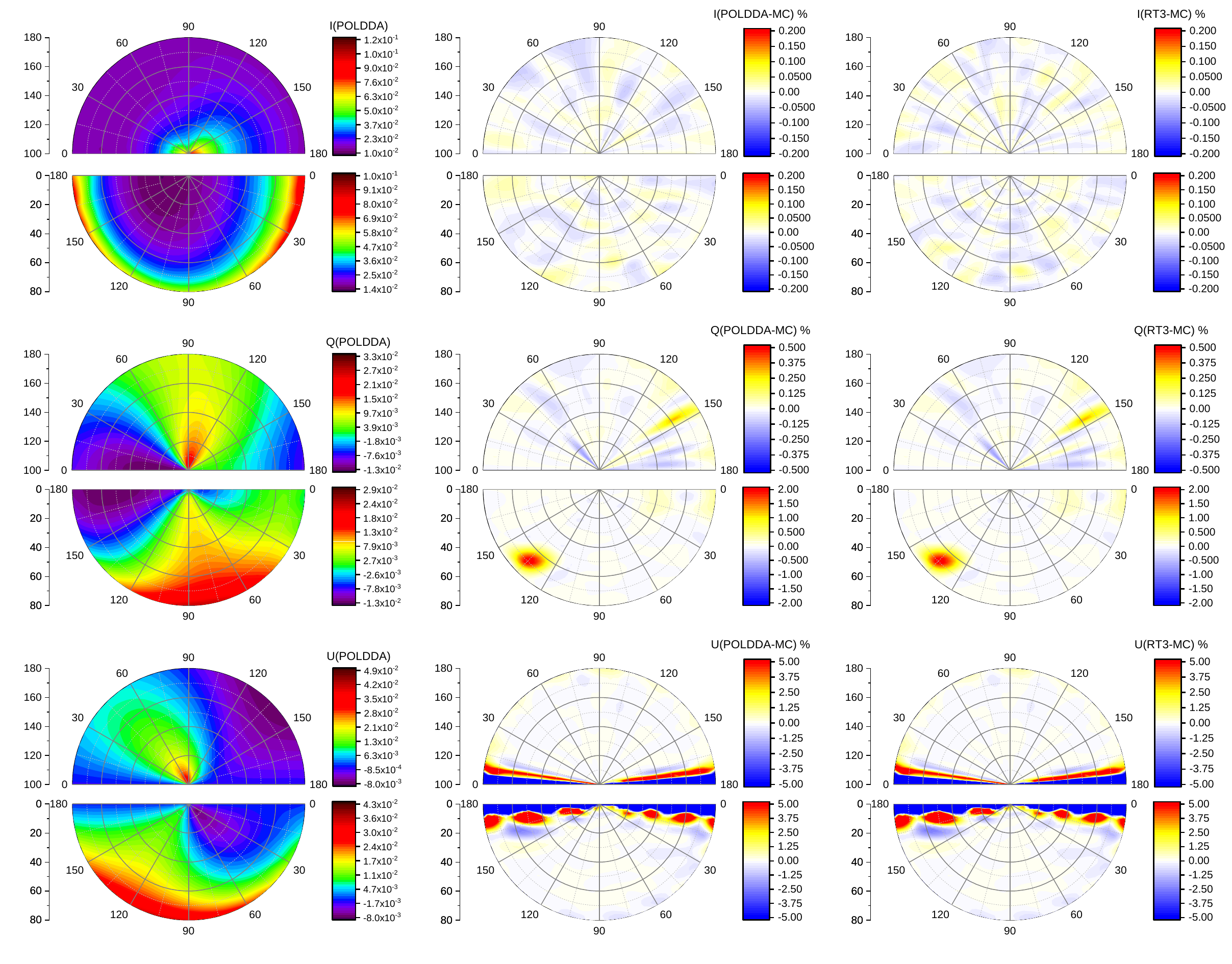}}
  \caption{Case 3: Multi-layer atmosphere with only Rayleigh scattering, (left column) the results for the $I$-component, $Q$-component, and $U$-component at the top of the layer based on POLDDA (16S), units: $Wm^{-2}\mu m^{-1}sr^{-1}$; (middle column) the relative differences of POLDDA (16S) against MYSTIC and (right column) the relative differences of RT3 (32S) against MYSTIC, units: \%.}
\end{figure}

\begin{figure}\label{t11}
  \centerline{\includegraphics[width=40pc,angle=0]{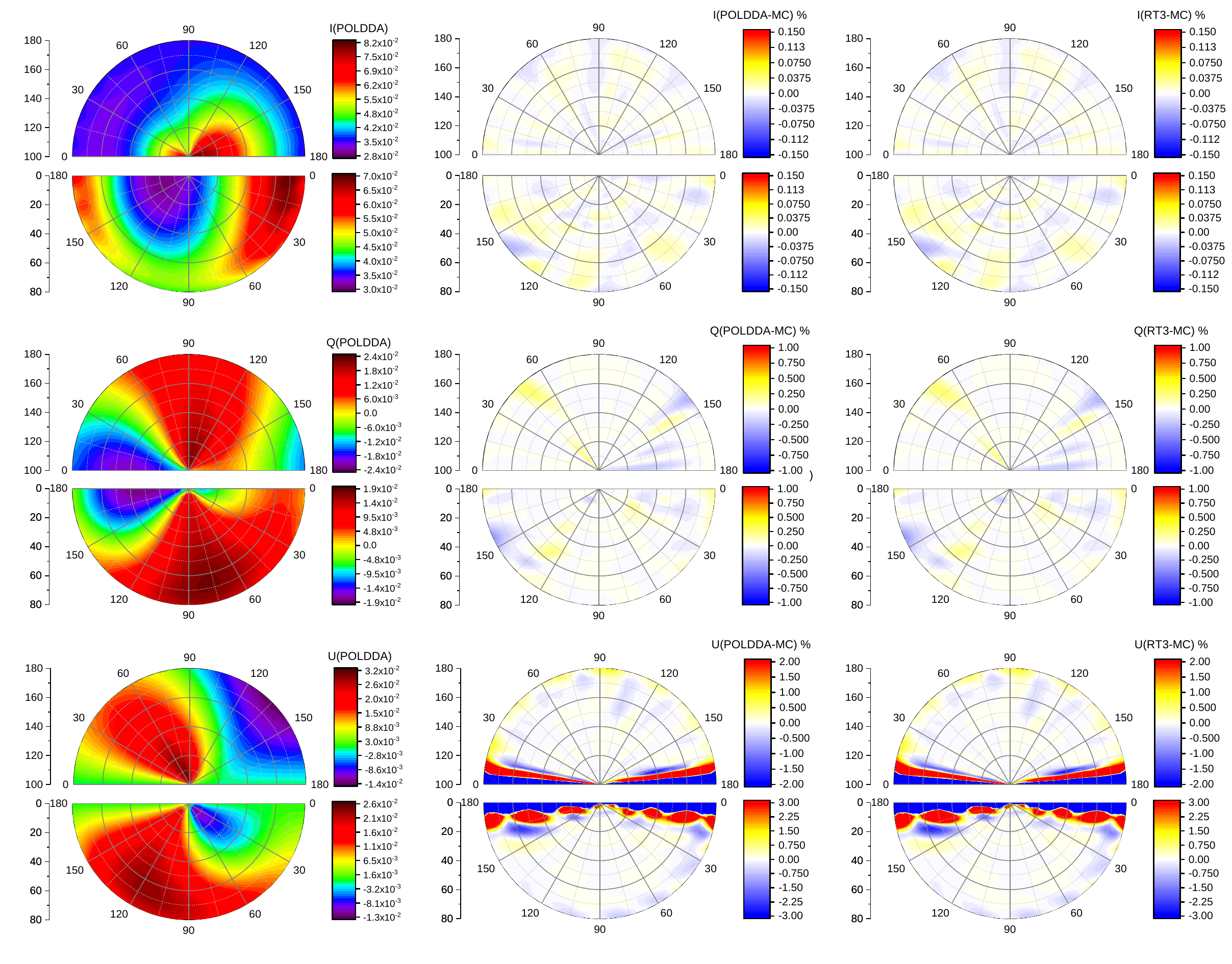}}
  \caption{Case 4: Multi-layer atmosphere with Rayleigh scattering and molecular absorption, (left column) the results for the $I$-component, $Q$-component, and $U$-component at the top of the layer based on POLDDA (16S), units: $Wm^{-2}\mu m^{-1}sr^{-1}$; (middle column) the relative differences of POLDDA (16S) against MYSTIC, and (right column) the relative differences of RT3 (16S) against MYSTIC, units: \%.}
\end{figure}

\begin{figure}\label{t12}
  \centerline{\includegraphics[width=36pc,angle=0]{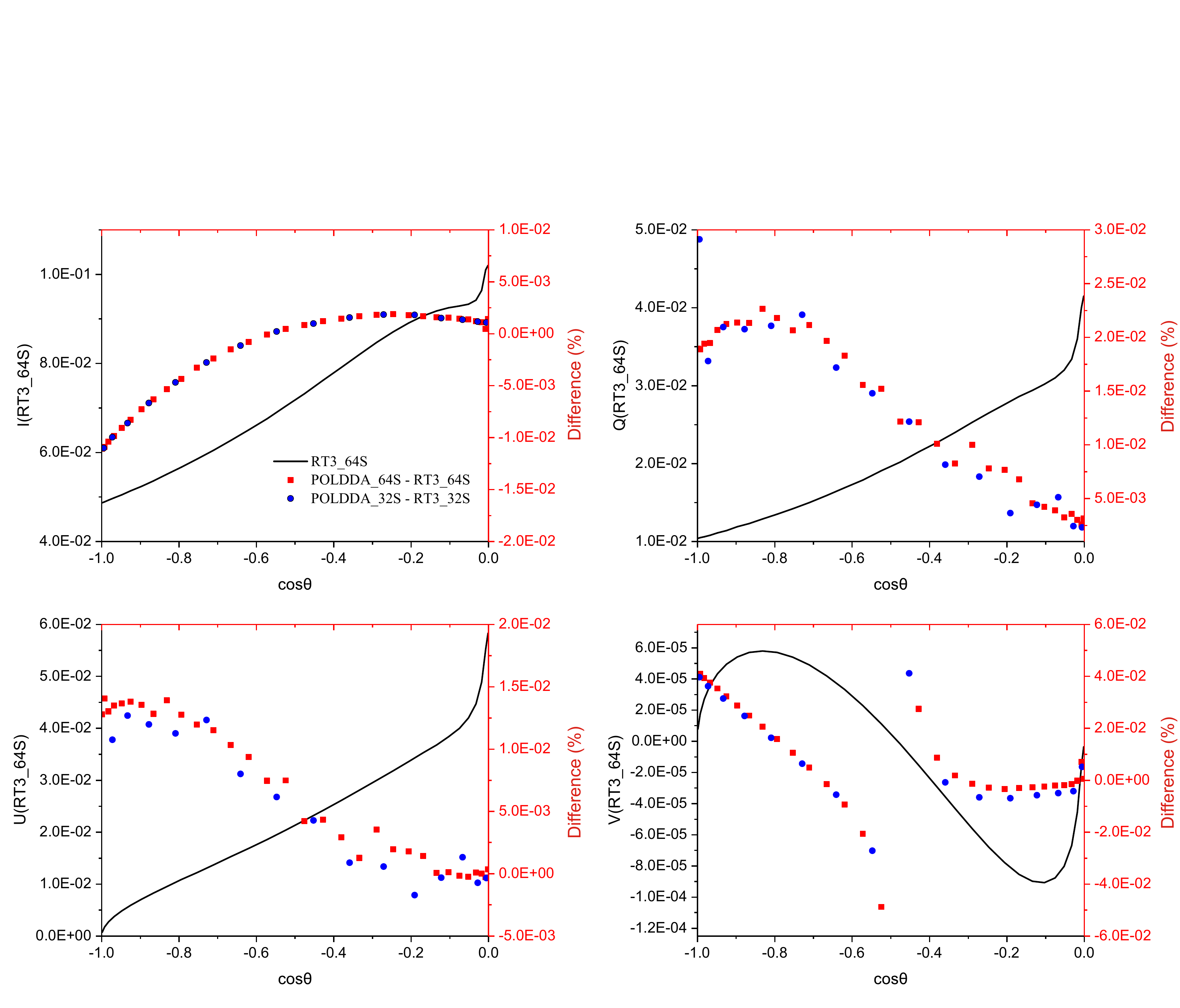}}
	\caption{Case 5: Multi-layer atmosphere with Rayleigh scattering and water cloud, the results for the $I$-component, $Q$-component, $U$-component, and $V$-component at the top of the layer based on RT3 (64S), units: $Wm^{-2}\mu m^{-1}sr^{-1}$, and the relative differences of POLDDA (64S) against RT3 (64S), POLDDA (32S) against RT3 (32S), units: \%.}
\end{figure}

\begin{figure}\label{t13}
  \centerline{\includegraphics[width=36pc,angle=0]{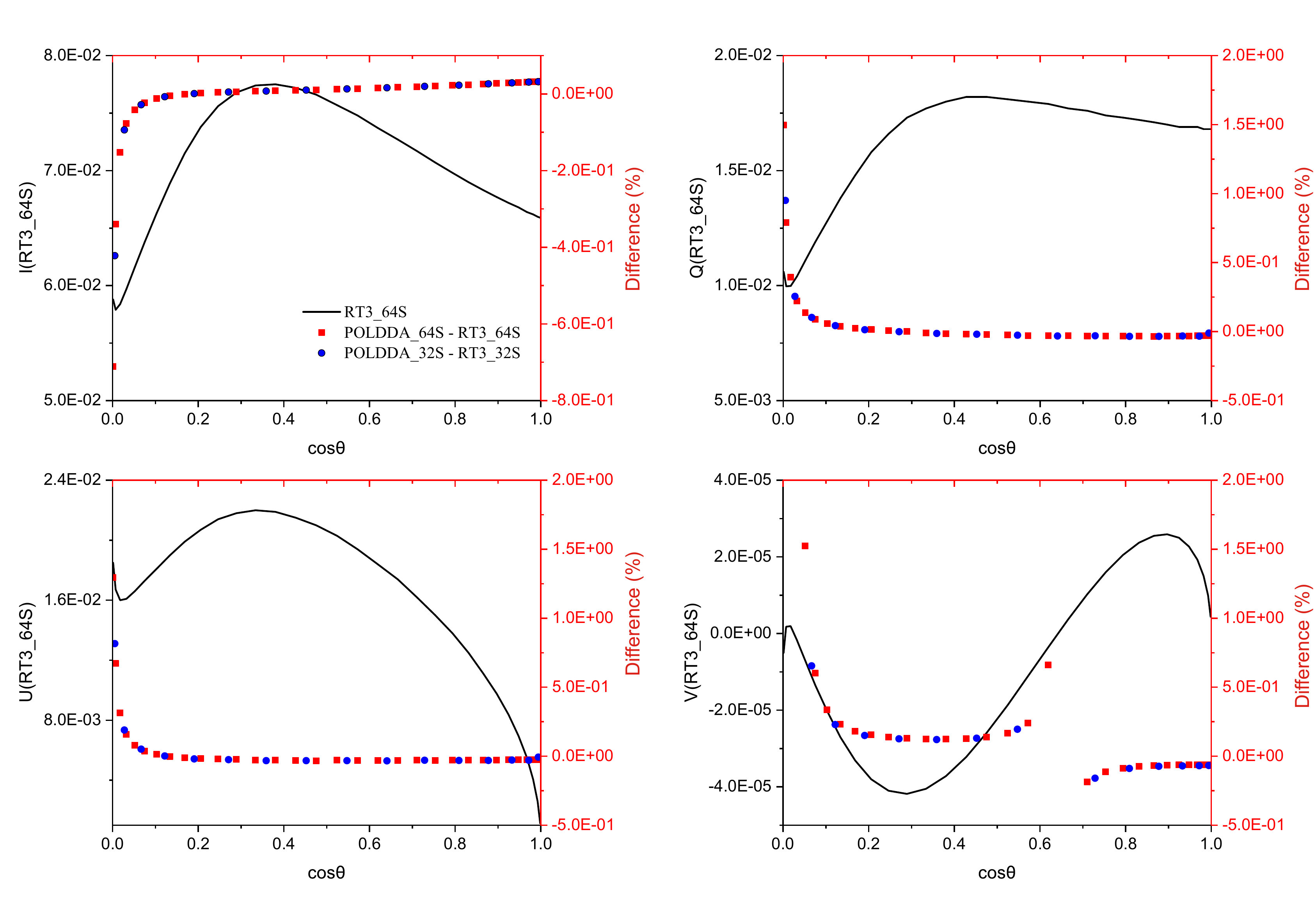}}
	\caption{Similar to Fig. 11 but for the results for the surface.}
\end{figure}

\begin{figure}\label{t14}
  \centerline{\includegraphics[width=36pc,angle=0]{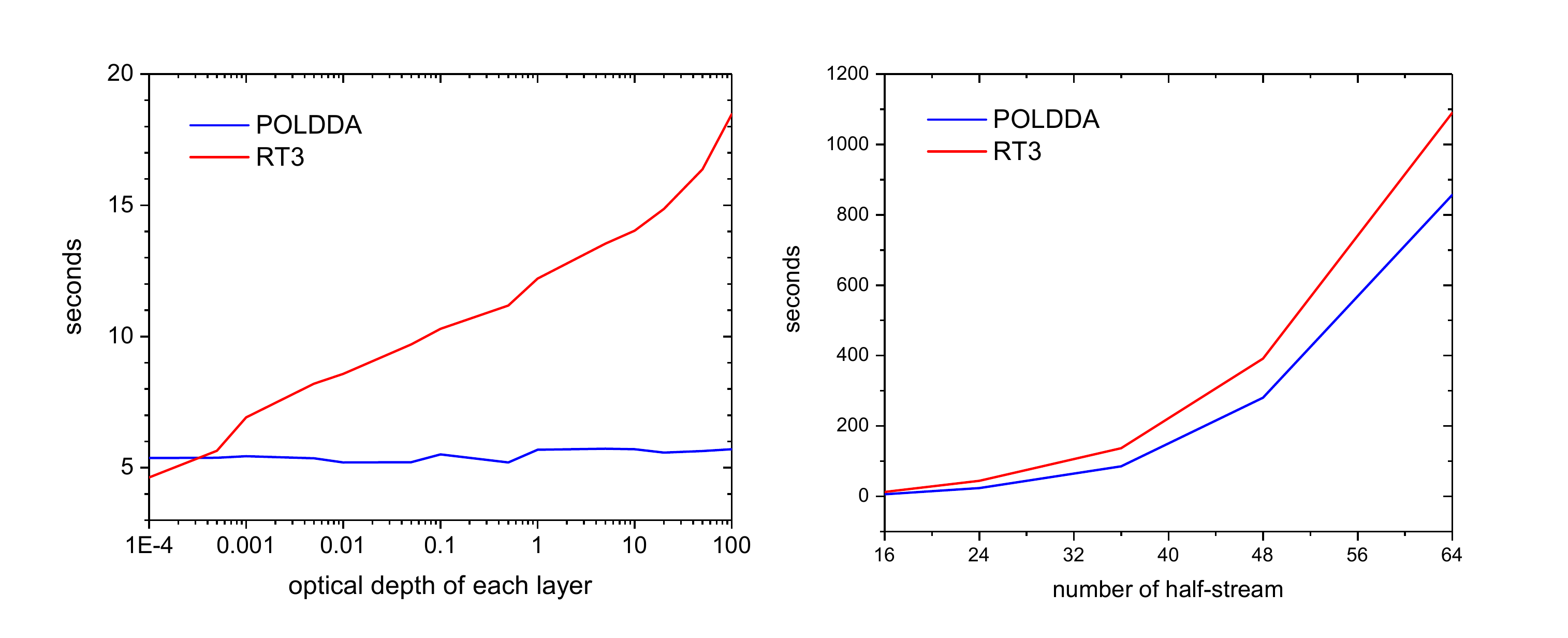}}
	\caption{Computational time (in seconds) of POLDDA and RT3 versus each layer, optical depth (left) and versus the number of streams (right).}
\end{figure}

\end{document}